\documentclass{article}

\usepackage{srcltx} 
\usepackage{epsfig}
\usepackage[T1]{fontenc}
\usepackage[utf8]{inputenc}
\usepackage[english]{babel}
\usepackage{times}
\usepackage{color}
\usepackage{amsmath,enumerate}
\usepackage{amssymb}
\usepackage{amscd}
\usepackage{latexsym}
\usepackage{enumerate}
\usepackage{tabularx}
\usepackage{mathrsfs}
\usepackage{url}
\usepackage{hyperref}
\usepackage[affil-it]{authblk}
\usepackage{easyReview}
\usepackage{gensymb}

\title{3-colorable planar graphs have an intersection segment representation using 3 slopes\thanks{This is the extended version of a paper presented at {\it WG '19}~\cite{G19}. This research is partially supported by the ANR GATO, under contract ANR-16-CE40-0009.}}

\author[a]{Daniel Gonçalves}

\affil[a]{{\small LIRMM, Université de Montpellier, CNRS, Montpellier, France.}%\\
}

\date{}

\newenvironment{proof}{\par \noindent \textbf{Proof.} }{\hfill$\Box$\medskip}

\newtheorem{theorem}{Theorem}

\newtheorem{definition}[theorem]{Definition}

\newtheorem{proposition}[theorem]{Proposition}
\newtheorem{lemma}[theorem]{Lemma}

\newtheorem{remark}[theorem]{Remark}
\newtheorem{claim}[theorem]{Claim}

\DeclareMathOperator{\R}{\mathcal R}
\DeclareMathOperator{\V}{\mathcal V}
\DeclareMathOperator{\Ls}{\mathcal L}

\begin{document}
\maketitle

\begin{abstract}
In his PhD Thesis E.R.~Scheinerman conjectured that planar graphs are
intersection graphs of line segments in the plane. This conjecture was
proved with two different approaches by J.~Chalopin and the author,
and by the author, L.~Isenmann, and C. Pennarun. In the case of
3-colorable planar graphs E.R.~Scheinerman conjectured that it is
possible to restrict the set of slopes used by the segments to only 3
slopes. Here we prove this conjecture by using an approach introduced
by S. Felsner to deal with contact representations of planar graphs
with homothetic triangles.
\end{abstract}

\section{Introduction}

In this paper, we consider intersection representations for planar graphs.  A
\emph{segment representation} of a graph $G$ maps every vertex $v\in V(G)$ to a
straight line segment ${\bf v}$ of the plane so that two segments ${\bf u}$ and ${\bf v}$ intersect if and only if $uv\in E(G)$.  
%Although this graph
% family is simply defined, it is not easy to manipulate.  Actually,
% even if this class of graphs is small (there are less than $2^{O(n\log
%   n)}$ such graphs with $n$ vertices~\cite{PS01}) a segment representation may
% be long to encode. Indeed, in the representations of some of these graphs, the endpoints
% of the segments need at least $2^{\sqrt{n}}$ bits to be
% coded~\cite{KM94}. There are also interesting open problems
% concerning this class of graphs.  For example, we know that deciding
% whether a graph $G$ admits a segment representation is NP-hard, actually it is
% even $\exists\mathbb{R}$-complete~\cite{K91}, but
% it is still open whether this problem belongs to NP or not. Here, we
% focus on segment representations for planar graphs.
%
%
In his PhD Thesis, E.R. Scheinerman~\cite{S84} conjectured that every
planar graph has a segment representation. This conjecture attracted
a lot of attention. H. de Fraysseix and P. Ossona de
Mendez~\cite{dFOdM07} proved it for a large family of planar graphs,
the planar graphs having a 4-coloring in which every induced cycle of
length 4 uses at most 3 colors. In particular, this implies the
conjecture for 3-colorable planar graphs. Then J.~Chalopin and the
author finally proved this conjecture~\cite{CG09}. More recently, a much
simpler proof was provided by the author, L.~Isenmann, and
C.~Pennarun~\cite{GIP18}. Here we focus on segment representations of
planar graphs with further restrictions.

\begin{figure}%[h!]
\begin{center}
\includegraphics[scale=0.3]{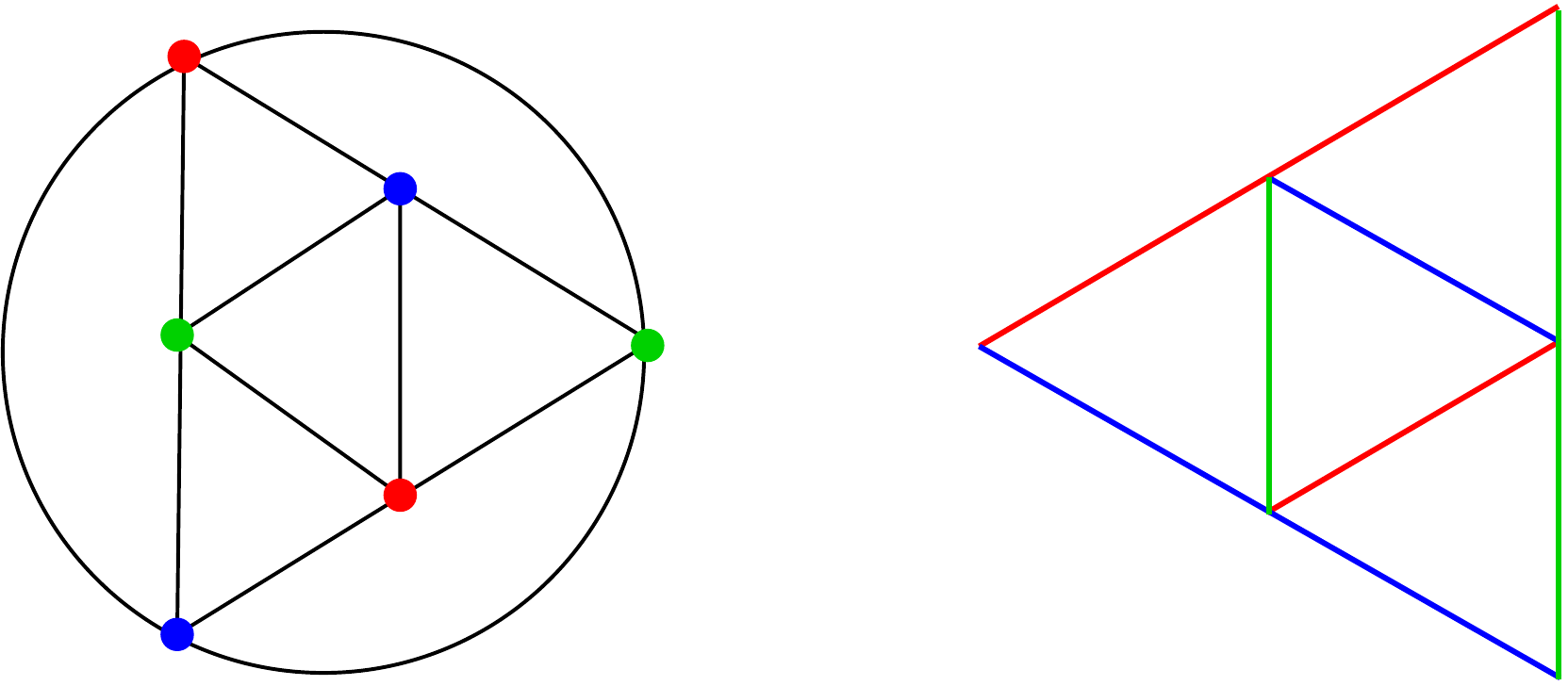}
\end{center}
\caption{The octahedron and a 3-slopes contact representation. It is unique, up to vertex automorphism, up to scaling, and once the slopes are set.}
\label{fig:octahedron}
\end{figure}

In his PhD Thesis, E.R. Scheinerman~\cite{S84} proved that every
outerplanar graph has a segment representation where only 3 slopes are
used, and where parallel segments do not intersect. Let us call such a
representation a \emph{3-slopes segment representation}.  This result
led E.R. Scheinerman conjecture~\cite{S-conj} (see
also~\cite{dFOdM07}) that such representation exists for every
3-colorable planar graph. Later, several groups proved a related
result on bipartite planar graphs~\cite{CKU98,FMP94,HNZ91}. They
proved that every bipartite planar graph has a 2-slopes segment
representation, with the extra property that segments do not cross
each other. Let us call such a representation a \emph{2-slopes contact
  segment representation}. Later, de Castro \emph{et
  al.}~\cite{CCDM02} considered a particular class of 3-colorable
planar graphs. They proved that every triangle-free planar graph has a
3-slopes contact segment representation. Such a contact segment
representation cannot be asked for any 3-colorable planar
graph. Indeed, up to isomorphism, the octahedron has only one 3-slopes
contact segment representation depicted in
Figure~\ref{fig:octahedron}, and one can easily check that this
representation does not extend to the (3-colorable) graph obtained
after gluing a copy of an octahedron in each of its faces.  However,
we will use 3-slopes contact segment representations in the proof of
our main result.
\begin{theorem}\label{thm:main}
Every 3-colored planar graph has a 3-slopes segment representation such
that parallel segments correspond to the color classes.
\end{theorem}

As such a representation implies 3-colorability, and as 3-colorability is NP-complete already for planar graphs~\cite{S73} we have that it is NP-hard to decide if a plane graphs has a 3-slopes segment representation in which no parallel segments intersect. Here, the NP-membership clearly follows from Theorem~\ref{thm:main}. The NP-membership is often non-trivial for segment representations. For example, we know that it is NP-hard to decide whether a graph $G$ admits a segment representation, actually it is even $\exists\mathbb{R}$-complete~\cite{K91}, but it is still open whether this problem belongs to NP or not.

It is easy to see that every 3-colored planar
graph is the induced subgraph of some 3-colored triangulation. Indeed, adding a new vertex in a non-triangular face, one can reduce the maximum length of such a face, or reduce the number of colors around such a face. 
As a 3-slopes segment representation of a triangulation naturally induces such a
representation for its induced subgraphs, we can restrict ourselves to the case of triangulations in the following.
In Section~\ref{sec:terminology} we provide some
basic definitions.  Section~\ref{sec:TC} is devoted to the
so-called \emph{triangular contact schemes}, which are particular 3-slopes contact segment representations of an auxiliary graph (to be formally defined there). It is shown that
every 3-colorable triangulation admits such a scheme. Then, those
schemes are used in Section~\ref{sec:3-dir} to build 3-slopes
segment representations.  Finally, we conclude with some remarks
on 4-slopes segment representations.

\section{Terminology}
\label{sec:terminology}

All graphs considered here are simple, that is without loops nor multiple edges.
A \emph{triangulation} is a plane graph where every face has size
three. A triangulation $T$
is \emph{Eulerian} if every vertex has even degree. It is well known
that these triangulations are the 3-colorable triangulations~\cite{Lov93}.
Actually, these triangulations are uniquely 3-colorable (up to color
permutation). Hence, their vertex set $V(T)$ is canonically partitioned
into three independent sets $A$, $B$ and $C$. In the following we will
denote the vertices of these sets respectively $a_i$ with $0\le i <
|A|$, $b_j$ with $0\le j < |B|$, and $c_k$ with $0\le k < |C|$.  In
such a triangulation $T$ any face is incident to one vertex $a_i$, one
vertex $b_j$, and one vertex $c_k$, and these vertices appear in this
order either clockwisely or counterclockwisely.  In the following, the
vertices of the outerface are always denoted by $a_0$, $b_0$ and $c_0$,
and they appear clockwisely in this order around $T$.  As the orders
of two adjacent faces are opposite, the dual graph of $T$ is
bipartite. Given a Eulerian triangulation $T$ with face set $F(T)$,
let us denote by $F_1(T)$ and $F_2(T)$ (or simply $F_1$ and $F_2$ if
it is clear from the context) the face sets partitioning $F(T)$, such
that no two adjacent faces belong to the same set, and such that
$F_2(T)$ contains the outer face. Note that by construction for any
face $f\in F_1(T)$ (resp. $f\in F_2(T)$) its vertices $a_i$, $b_j$ and
$c_k$ appear in clockwise (resp. counterclockwise) order around
$f$. Note that the vertices $a_0$, $b_0$ and $c_0$ appear in clockwise
order around $T$, but in counterclockwise order with respect to the outer
face. Let $n=|V(T)|$. As $T$ is a triangulation, by Euler's formula it
has $2n-4$ faces. Hence, as $T$'s dual is bipartite and 3-regular,
$|F_1(T)| = |F_2(T)| = n-2$.

In the following, we build 3-slopes segment representations. The 3
slopes used are expected to be distinct, but apart from that the exact
3 slopes considered do not matter. Indeed, for any two triples of
slopes, $(s_1,s_2,s_3)$ and $(s'_1,s'_2,s'_3)$, there exists an affine
map of the plane turning any 3-slopes segment representation using
slopes $(s_1,s_2,s_3)$ into a 3-slopes segment representation using
slopes $(s'_1,s'_2,s'_3)$.  We denote by $\overrightarrow{a}$,
$\overrightarrow{b}$, and $\overrightarrow{c}$ the vectors
corresponding to slopes of the sets $A$, $B$ , and $C$
respectively. The magnitude of these vectors is chosen such that
$\overrightarrow{a} + \overrightarrow{b} + \overrightarrow{c}
= \overrightarrow{0}$.

\section{TC-representations and TC-schemes}
\label{sec:TC}

\begin{figure}%[h!]
\begin{center}
    \scalebox{0.32}{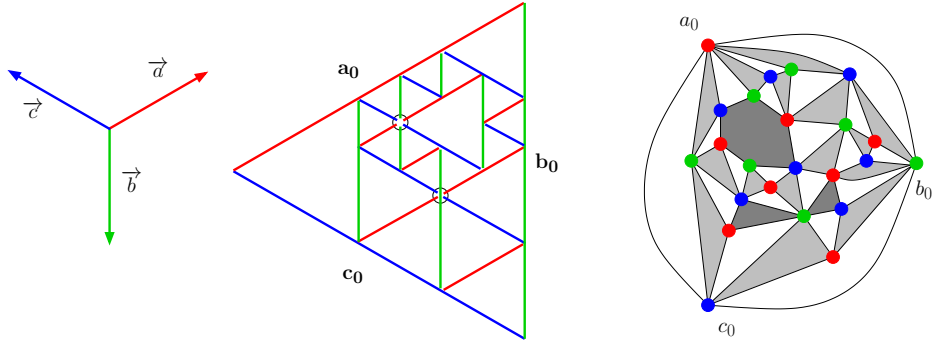}
\end{center}
\caption{(left) Vectors $\overrightarrow{a}$, $\overrightarrow{b}$,
  and $\overrightarrow{c}$ (middle) A TC-representation $\R$ with various
  types of intersection points. The circled ones correspond to particular intersection points with more than three segments. (right) Its corner graph $C(\R)$, where gray
  faces correspond to the degenerate faces (i.e. they correspond to intersection points in $\R$). The dark gray faces are particular degenerate faces. One has size six, and there
  are two faces of size three that correspond to the same intersection
  point.}
\label{fig:TC-graph}
\end{figure}

We begin with the definition of the particular type of 3-slopes contact
representations illustrated in Figure~\ref{fig:TC-graph}.
\begin{definition}
For the fixed vectors $\overrightarrow{a}$,
$\overrightarrow{b}$, $\overrightarrow{c}$ as defined previously, a \emph{Triangular 3-slopes Contact segment representation}
(\emph{TC-representation} for short) is a 3-slopes contact segment
representation using the same slopes as $\overrightarrow{a}$,
$\overrightarrow{b}$, and $\overrightarrow{c}$, and where:
\begin{itemize}
\item Three segments ${\bf a_0}$, ${\bf b_0}$, and ${\bf c_0}$, 
form a triangle which contains all the other segments. 
Going clockwise around this outer-triangle, one successively follows $\alpha\overrightarrow{a}$,
$\alpha\overrightarrow{b}$, and then $\alpha\overrightarrow{c}$, with $\alpha=1$.
\item Every inner region is a triangle, where each side is contained in a
  segment of the representation.
\item Two parallel segments intersect in at most one point, and this point is an endpoint of both segments.
\end{itemize}
\end{definition}

\begin{remark}
In such a representation, an intersection point ${\bf p}$ is of one of
the following four types (see Figure~\ref{fig:TC-graph}).
\begin{itemize}
\item The intersection point of 2 outer segments;
\item the intersection point of 3 segments (with 3 distinct slopes)
  such that exactly two of them end at ${\bf p}$;
\item the intersection point of 5 segments such that exactly four of
  them end at ${\bf p}$ (such a point will be generally considered as
  the merge of two intersection points of 3 segments); or
\item the intersection point of 6 segments that have an end at ${\bf p}$.
\end{itemize}
\end{remark}

\begin{definition}
Given a TC-representation $\R$, let the \emph{corner graph} $C(\R)$ be the plane graph whose vertices correspond to the segments of the
representation, and where two vertices are adjacent if and only if the
corresponding segments form a corner of one of the inner
triangles. The orders of the neighbors around a vertex $v$ correspond
to the order of the segments around the segment ${\bf v}$.
\end{definition}

Note that the corner graph of a TC-representation has several
properties. For example, since two parallel segments cannot form a corner of a triangle, they correspond to non-adjacent
vertices. The slopes hence define a 3-coloring of the graph.  Note
also that the dual graph of $C(\R)$ is bipartite. Indeed, such a plane graph
has two types of faces, one set contains the (triangular) faces
corresponding to the inner regions of the TC-representation, and the
other set contains the outerface and the faces corresponding to
intersection points.  Let us denote the latter faces \emph{degenerate
  faces}, and note that those faces have size three or six. A size six
face $(a_i,b_j,c_k,a_{i'},b_{j'}, c_{k'})$ comes from the intersection
point of six segments, and as those six segments go in distinct
directions they do not intersect elsewhere, so this cycle has no chord
in $C(\R)$. Note that going clockwise on the border of any
inner region of $\R$, since $\overrightarrow{a} + \overrightarrow{b} + \overrightarrow{c} = \overrightarrow{0}$, one successively follows $\alpha\overrightarrow{a}$,
$\alpha\overrightarrow{b}$, and then $\alpha\overrightarrow{c}$, for some not necessarily positive value $\alpha$. This value $\alpha$ is the \emph{size} of this inner triangle. Note that the outer-triangle formed by ${\bf a_0}$, ${\bf b_0}$ and ${\bf c_0}$ has size 1. In our illustrations, the triangles of positive size have a corner on the left.
Let us now prove that any corner graph is 3-connected.
\begin{lemma}\label{lem:TC-3-connected}
    For any TC-representation $\R$, its corner graph $C(\R)$ is 3-connected, and hence it has only one planar embedding with $a_0,b_0,c_0$ on the outerboundary in clockwise order.
\end{lemma}
\begin{proof}
Towards a contradiction, suppose $C(\R)$ is not 3-connected for some TC-representation $\R$, and let $\{x,y\}$ be a separating set. The case where there is a vertex cut but no separating set of size two is trivial.
We assume without loss of generality that $x$ is a $A$-vertex, then by symmetry we just have to consider the case where $y$ is a $A$-vertex, and the case where $y$ is a C-vertex.

If both $x$ and $y$ are $A$-vertices, let us prove that any connected component $H$ of $C(\R)\setminus \{x,y\}$ contains $b_0$. Indeed, let ${\bf b_j}$ be the rightmost vertical segment of $H$. If $b_j\neq b_0$ there is a $C$-vertex $c_k$ such that ${\bf c_k}$ intersects ${\bf b_j}$ on its topmost point and such that ${\bf c_k}$ continues further on the right. Indeed, the topmost inner triangle incident to the right of ${\bf b_j}$ is incident to such segment ${\bf c_k}$. Then the rightmost point of ${\bf c_k}$ has to belong to some vertical segment on the right of $b_j$, a contradiction.

If $x$ is a $A$-vertex and $y$ is a $C$-vertex, we have to distinguish two subcases according to how they intersect (or not).

Suppose first that ${\bf x}$ and ${\bf y}$ are not on the boundary of an inner triangle, of negative size (i.e. pointing to the right). In this case, we can proceed as in the previous case by considering the rightmost vertical segment of any connected component $H$ of $C(\R)\setminus \{x,y\}$. If $b_j\neq b_0$, the segment ${\bf b_j}$ has (at least) $2t$ incident segments going further to the right, where $t\ge 1$ is the number of inner faces incident to ${\bf b_j}$, on its right. If some of these $2t$ neighbors in $C(\R)$ is distinct from $x$ and $y$, this segment brings us to another $B$-vertex on the right, as in the previous case. If these $2t$ neighbors correspond exactly to $x$ and $y$, we have $t=1$ and we are in the case where ${\bf x}$ and ${\bf y}$ are on the boundary of an inner triangle, of negative size, a contradiction.

Suppose now that ${\bf x}$ and ${\bf y}$ are on the boundary of an inner triangle, of negative size (i.e. pointing to the right). In that case, we are going to prove that any connected component $H$ of $C(\R)\setminus \{x,y\}$ contains the vertex $b_t$, corresponding to the leftmost vertical segment; the one forming an inner triangle with ${\bf a_0}$ and ${\bf c_0}$. We proceed similarly as above by letting ${\bf b_j}$ be the leftmost vertical segment of $H$, and by considering its (at least) $2t$ incident segments going further to the left, where $t\ge 1$ is the number of inner faces incident to ${\bf b_j}$, on its left. 
If some of these neighbors in $C(\R)$ is distinct from $x$ and $y$, this segment would bring us to another $B$-vertex on the left, a contradiction. We are thus left with the case where $t=1$ and where there are only two such neighbors, which would be $x$ and $y$. This is impossible because in that case ${\bf x}$ and ${\bf y}$ would be on the boundary of an inner triangle, of positive size, and it is impossible in a TC-representation for two segments to have more than one inner triangle in common.
\end{proof}

\begin{figure}%[h!]
\begin{center}
\includegraphics[scale=0.18]{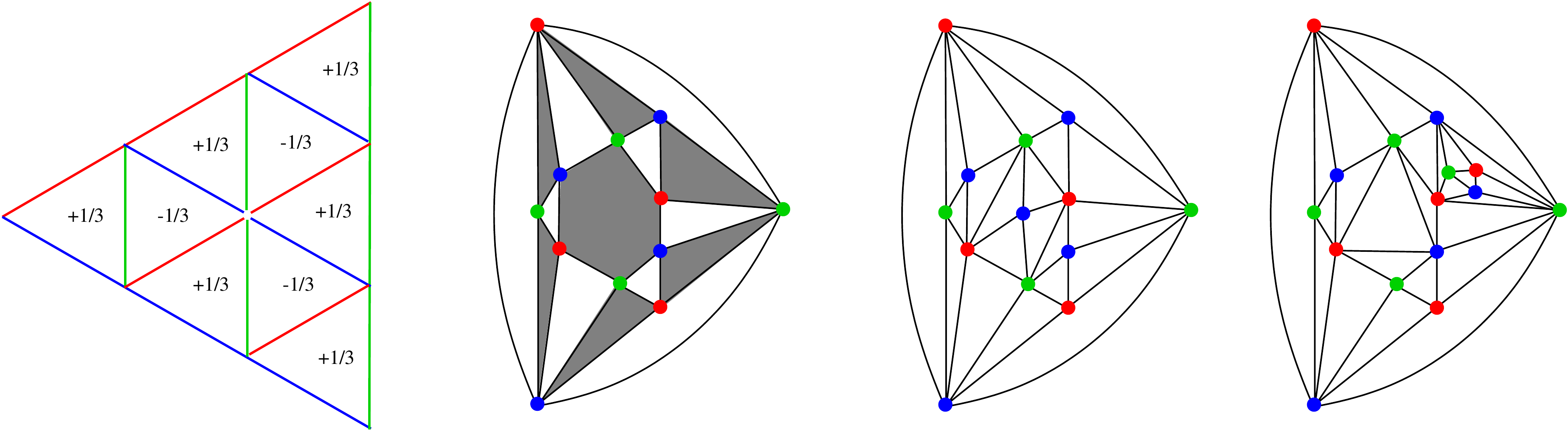}
\end{center}
\caption{From left to right. A TC-representation $\R$, with the size of its inner triangles; its corner graph
  $C(\R)$, where gray faces are the degenerate faces; and two
  triangulations where $\R$ is a TC-scheme of both.}
\label{fig:TC-graph-2}
\end{figure}

\begin{lemma}\label{lem:equiv-def-scheme}
  Consider  a Eulerian triangulation $T$ and a TC-representation $\R$ such that each segment of $\R$ corresponds to a distinct vertex of $T$.
  The following two statements are equivalent:
    \begin{enumerate}
        \item[(1)] $C(\R)$ is a submap of $T$ with the same
outer face as $T$, such that the vertices and edges of
$T$ not in $C(\R)$ lie inside degenerate faces of $C(\R)$.
        \item[(2)] The outerface and the non-degenerate faces of $C(\R)$ are, respectively, the outerface and a subset of the inner faces of $T$.
    \end{enumerate}
\end{lemma}
If (1) (or equivalently (2)) is fulfilled, we call $\R$ a \emph{TC-scheme} of  $T$ (see Figure~\ref{fig:TC-graph-2}).
\begin{proof}
In both cases, the outerfaces of $C(\R)$ and $T$ are asked to be the same, so we can focus on the inner faces.
    \paragraph{(1) $\Longrightarrow$ (2).}
    If $C(\R)$ and $T$ fulfill (1), then any non-degenerate face of $C(\R)$ induces a cycle in $T$ (as $C(\R)$ is a subgraph of $T$). Furthermore, this cycle does not contain any edge or vertex of $T$. It thus bounds an inner face of $T$. 

    \paragraph{(2) $\Longrightarrow$ (1).}
    In a corner graph $C(\R)$, every edge belongs to a non-degenerate inner-face (by definition). Hence, if $C(\R)$ and $T$ fulfill (2), any edge of $C(\R)$ is an edge of $T$. $C(\R)$ is thus a subgraph of $T$. Furthermore, as they have same outerface, and as $C(\R)$ has only one possible embedding (by Lemma~\ref{lem:TC-3-connected}) we have that $C(\R)$ is a submap of $T$. The non-degenerate faces of $C(\R)$ being inner faces of $T$ (by (2)) the vertices and edges of $T$ not in $C(\R)$ have to lie inside the degenerate faces of $C(\R)$.
\end{proof}

% \begin{definition}
% A TC-representation $\R$ is a \emph{TC-scheme} of a Eulerian
% triangulation $T$ if $C(\R)$ is a submap of $T$ with the same
% outer face as $T$, and such that the vertices and edges of
% $T$ not in $C(\R)$ lie inside degenerate faces of $C(\R)$ (see
% Figure~\ref{fig:TC-graph-2}).
% \end{definition}
% Actually, as in $C(\R)$, the inner faces around any vertex alternate
% among degenerate and non-degenerate. This implies that every edge of
% $C(\R)$ bounds a non-degenerate face, and a face that is degenerate or that
% is the outerface. We thus have the following.
% \begin{remark}~\label{rmk:TC-scheme}
% A TC-representation  $\R$ is a TC-scheme of $T$ if and only if the
% non-degenerate faces of $C(\R)$ and its outerface are faces of $T$.
% \end{remark}
The main ingredient in the proof of Theorem~\ref{thm:main} is the following.
\begin{theorem}\label{thm:TC}
Every Eulerian triangulation $T$ has a TC-scheme, and this scheme is
unique.
\end{theorem}

To prove this theorem we use TC-schemes as defined in Lemma~\ref{lem:equiv-def-scheme}.(2). The definition in Lemma~\ref{lem:equiv-def-scheme}.(1) will be used in Section~\ref{sec:3-dir} to complete the proof of Theorem~\ref{thm:main}.
To prove Theorem~\ref{thm:TC}, we first
model TC-schemes of $T$ by means of a system of linear equations in
Subsection~\ref{ssec:TC-model}. We then show in Subsection~\ref{ssec:TC-sol}
that such a linear system always has a solution, and that this
solution is unique (c.f. Lemma~\ref{lem:det}).  Finally, we show in
Subsection~\ref{ssec:TC-bij} that the solution of this linear system
defines a TC-scheme of $T$ (c.f. Lemma~\ref{lem:TC-scheme}).

\subsection{The linear system model}\label{ssec:TC-model}

In a TC-representation all the triangles are homothetic, and they are included in the outer triangle which has size one. Note that we will have positive, negative, and degenerate
(zero) homothets. Recall that 
the size of a triangle is its relative size with respect to the
outer triangle.  The
variables of our linear system correspond to the sizes of the
triangular regions. So for each face $f\in F_1$ we have a
variable $x_f$. Informally, the value of $x_f$ will prescribe the size
and shape of the corresponding triangle in a TC-representation. If
$x_f<0$, $x_f=0$, or if $x_f>0$ the corresponding triangle has a
corner on the right, is missing, or has a corner on the left,
respectively.

\begin{figure}%[h!]
\begin{center}
\includegraphics[scale=0.4]{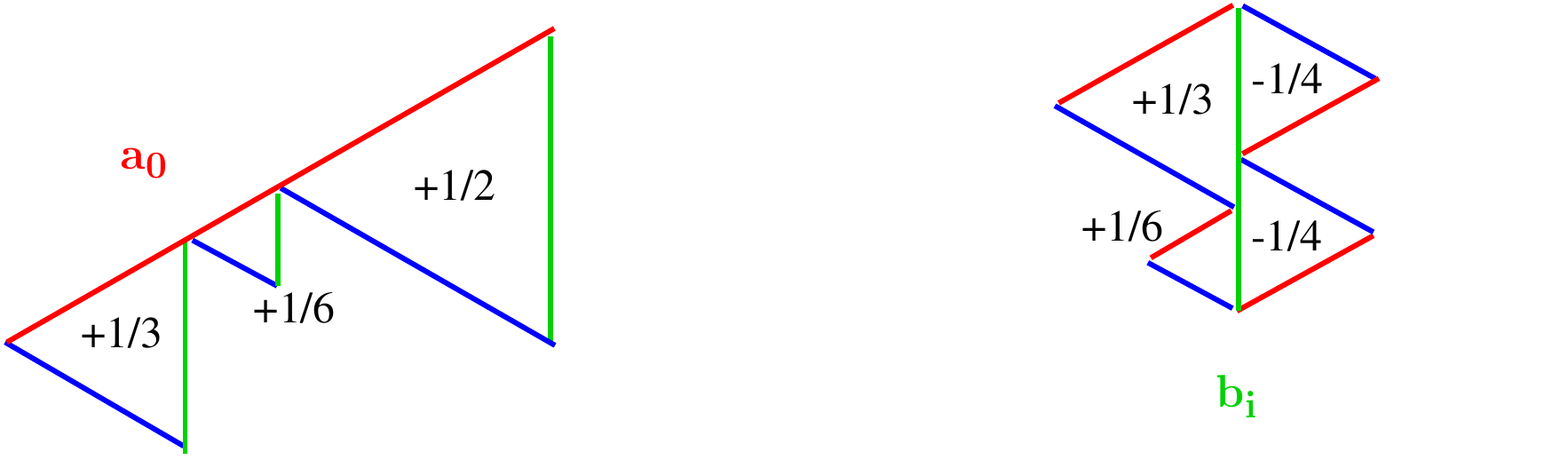}
\end{center}
\caption{(left) The size of the triangles around $a_0$. (right) The size of the triangles around some inner vertex $b_i$.}
\label{fig:systemLE}
\end{figure}

Let us denote by $F_1(v)$ the subset of faces of $F_1$ incident to
$v$. As the outer triangle has size 1 and contains the other
triangles, the faces in $F_1(a_0)$ should have non-negative sizes, and
they should sum up to 1 (see Figure~\ref{fig:systemLE}, left). We hence
consider the following constraint.
\begin{equation}
  \sum_{f\in F_1(a_0)} x_f =1\tag{$a_0$}\label{eqn:a0}
\end{equation}
We add no constraint about the sign of these sizes. Note that
similar constraints hold for $b_0$ and $c_0$.
\begin{equation}
  \sum_{f\in F_1(b_0)} x_f =1\tag{$b_0$}\label{eqn:b0}
\end{equation}
\begin{equation}
  \sum_{f\in F_1(c_0)} x_f =1\tag{$c_0$}\label{eqn:c0}
\end{equation}

Similarly, around an inner segment of a TC-representation all the
triangles on one side have the same size sign, which is opposite to the
other side (see Figure~\ref{fig:systemLE}, right). Furthermore, by summing all of these sizes one should obtain 0. Hence, for any inner vertex
$u$ we consider the following constraint.
\begin{equation}
  \sum_{f\in F_1(u)} x_f =0\tag{$u$}\label{eqn:u}
\end{equation}

In the following, Equation~($a_j$) will refer to
Equation~\eqref{eqn:u} where vertex $u$ is replaced by $a_j$.  Note
that every face $f\in F_1$ is incident to exactly one vertex of $A$,
one vertex of $B$, and one vertex of $C$.  Hence by summing
Equations~\eqref{eqn:a0}, ($a_1$),\ldots,($a_{|A|}$), one obtains that
$\sum_{f\in F_1} x_f =1$.
%% \begin{equation}
%%   \sum_{f\in F_1} x_f =1\tag{$\forall f$}\label{eqn:f}
%% \end{equation}
The same holds with Equations~\eqref{eqn:b0},
($b_1$),\ldots,($b_{|B|}$), or with Equations~\eqref{eqn:c0},
($c_1$),\ldots,($c_{|C|}$). Equations~\eqref{eqn:b0} and
\eqref{eqn:c0} are hence implied by the others, and thus we do not need
to consider them anymore. Let us denote by $\Ls$ the obtained system
of $n-2$ linear equations on $|F_1|=n-2$ variables.

\subsection{$\Ls$ has a unique solution}\label{ssec:TC-sol}

Let us define the set $V' = V\setminus\{b_0,c_0\}$ of size $n-2$. 
Finding a solution to $\Ls$ is equivalent to finding a vector $S\in
\mathbb{R}^{F_1}$ (that is a vector indexed by elements of $F_1$) such
that $MS=I$, where $M\in \mathbb{R}^{V' \times F_1}$ (a square matrix
indexed by elements of $V' \times F_1$) and $I\in \mathbb{R}^{V'}$ are
defined by
\begin{eqnarray*}
  \label{eq:M}
M(x_i,f)=\left\{
\begin{array}{l}
1\ \ \ \mbox{ if } f\in F_1(x_i)\\
0\ \ \ \mbox{ otherwise. }
\end{array}\right.
& \ \ \ \ \
&
I(x_i)=\left\{
\begin{array}{l}
1\ \ \ \mbox{ if } x_i=a_0\\
0\ \ \ \mbox{ otherwise. }
\end{array}\right.
\end{eqnarray*}

Given some bijective mappings $g_{V'} : [1,\ldots,n-2] \longrightarrow
V'$ and $g_{F_1} : [1,\ldots,n-2]\longrightarrow F_1$, one can index
the elements of $M$ by pairs $(i,j) \in [1,\ldots,n-2]
\times [1,\ldots,n-2]$, and thus define the determinant of $M$.  By
the following lemma, $\Ls$ has a solution vector $S$, and this
solution is unique.
\begin{lemma}\label{lem:det}
The matrix $M$ is non-degenerate, i.e. $\det(M) \neq 0$.
\end{lemma}
The proof of this lemma is inspired by the work of S. Felsner~\cite{F09}
on contact representations with homothetic triangles. See also~\cite{FSS18}
for another proof using the same approach.
\begin{proof}
Let $T_M$ be the bipartite graph with independent sets $V'$ and $F_1$
such that $v\in V'$ and $f\in F_1$ are adjacent if and only if $v$ and
$f$ are incident in $T$.  Note that $M$ is the biadjacency matrix of
$T_M$.  From the embedding of $T$ one can easily embed $T_M$ in such a
way that all the inner faces have size 6, and such that $a_0$ is on
the outerboundary.

Note that every perfect matching of $T_M$ (if any) corresponds to a
permutation $\sigma$ on $[1,\ldots,n-2]$
%% (we say $\sigma$ belongs to the permutation group $S_{n-2}$)
defined by $\sigma(g_{V'}^{-1}(v))=g_{F_1}^{-1}(f)$, for any edge $vf$
of the perfect matching. If the obtained permutation is even we call
such perfect matching positive, otherwise it is negative.  From the
Leibniz formula for the determinant,
$$ \det(M) = \sum_{\sigma\in S_{n-2}} sgn(\sigma) \prod_{i\in
  [1,\ldots,n-2]} M(g_{V'}(i),g_{F_1}(\sigma(i)))
$$ one can see that $\det(M)$ counts the number of positive perfect
matchings of $T_M$ minus its number of negative perfect matchings.

\begin{claim}\label{cl:m-exists}
  The graph $T_M$ admits at least one perfect matching.
%% For every Eulerian triangulation $T$ there exists a bijection $\beta$
%% from $V'$ to $F_1$ such that for all $v\in V'$, $v$ is incident to
%% $\beta(v)$. Hence, the graph $T_M$ admits at least one perfect
%% matching.
\end{claim}
\begin{proof}
  As $T_M$ is bipartite, and as $|V'|=|F_1|$, it suffices to show that
  $T_M$ has an $F_1$-saturating matching.  This follows from Hall's
  marriage theorem, and the fact that for any set $X\subseteq F_1$ the
  set $Y\subset V'$ of vertices incident to a face in $X$ is such that
  $|Y|\ge |X|$. Let us show this below for any set $X\subseteq F_1$.

  Consider the (planar) subgraph of $T$ with all the edges and all the
  vertices incident to a face of $X$. Then, triangulate this graph and
  denote by $T_X$ the obtained triangulation. Note that as any two faces
  of $X$ are not adjacent in $T_X$, this triangulation has at least
  $2|X|$ faces. Indeed, around each vertex there are at least twice as
  many faces as faces of $X$, and summing over every vertex one
  obtains the inequality. Together with the fact that $T_X$ has
  $2|V(T_X)|-4$ faces,
$$ 2|V(T_X)|-4 \ge 2|X|$$
  and that $V(T_X) \subseteq Y\cup \{b_0,c_0\}$,
  $$|Y|+2 \ge |V(T_X)|$$
  one obtains that
$$ 2|V(T_X)|-4 +2|Y|+4\ge 2|X|+2|V(T_X)|$$
  $$|Y|\ge |X|$$
\end{proof}

Given a graph $G$ and a perfect matching $M$ of $G$, an
\emph{alternating cycle} $C$ is a cycle of $G$ with edges alternating
between $M$ and $E(G)\setminus M$. Note that replacing in $M$ the
edges of $M\cap C$ by the edges of $C\setminus M$ yields another
perfect matching. We call such operation a \emph{cycle exchange}.  It
is folklore that the perfect matchings of a graph are linked by
cycle exchanges. Indeed, given any perfect matching $M_1$ of $G$ one
can reach any perfect matching $M_2$, by a succession of cycle
exchanges. Actually, for $T_M$ any such cycle has length congruent to
$2\ (\bmod 4)$.

\begin{claim}\label{cl:m-same-sign}
For any perfect matching of $T_M$ and any of its alternating cycles
$C$, we have that the length $\ell(C)$ of $C$ is congruent to $2\ (\bmod 4)$.
\end{claim}
\begin{proof}
The subgraph $G$ of $T_M$ induced by the vertices and edges on or
inside $C$ is such that all the inner faces have length 6, and it is
routine from Euler's formula to show that $C$ has length congruent to
$2 + 2|V_G|\ (\bmod 4)$, where $V_G$ is the vertex set of $G$. Indeed,
$$ \ell(C)-6 + 6|F_G|= \sum_{f\in F_G} \ell(f) = 2|E_G| = 2|V_G| +2|F_G| -4$$
$$ \ell(C) \equiv 2|V_G| +2\ (\bmod 4)$$
Finally, as the vertices of $G$ are paired by the perfect matching we have
that $|V_G|$ is even.
\end{proof}

The previous claim implies that all the perfect matchings of $T_M$
induce permutations of the same sign. Indeed, performing a $(4k+2)$-cycle
exchange does not change the sign of the permutation, as it corresponds
to performing $2k$ transpositions in the permutation. Hence, all the
terms of $\det(M)$ have the same sign, and this sum has at least one
non-zero term (by Claim~\ref{cl:m-exists}). Thus $\det(M)\neq 0$.
\end{proof}

\subsection{A solution of $\Ls$ defines a TC-scheme}\label{ssec:TC-bij}

A TC-scheme $\R$ \emph{corresponds} to a solution of $\Ls$, the linear
system defined for a Eulerian triangulation $T$, if $\R$ is a
TC-scheme of $T$ such that each face $f\in F_1$ corresponds to an
inner triangle of $\R$ of size $x_f$, the solution of $\Ls$, unless
$x_f=0$. In other words, the inner regions of $\R$ correspond to
non-zero valued faces of $F_1$.

Let us now proceed to the main result of this section.
\begin{lemma}\label{lem:TC-scheme}
  Every Eulerian triangulation $T$ admits a TC-scheme $\R$ that
  corresponds to the solution of its linear system $\Ls$.
\end{lemma}
\begin{proof}
  We proceed by induction on the number of faces $f\in F_1$ such
  that $x_f=0$. We start with the case where every face $f\in F_1$ is
  such that $x_f\neq 0$.

\paragraph{If every face $f\in F_1$ is such that $x_f\neq 0$,} we construct 
a TC-scheme $\R$ corresponding to the solution of $\Ls$ as
follows. First let $\Delta$ be a triangle formed by three vectors
$\overrightarrow{a}$, $\overrightarrow{b}$, and $\overrightarrow{c}$
in this order. The sides of $\Delta$ correspond to ${\bf a_0}$, ${\bf
  b_0}$, and ${\bf c_0}$, respectively. For each face $f\in F_1$, let
$\Delta_f$ be a homothetic copy of $\Delta$ with ratio $x_f$. The
triangle $\Delta_f$ is thus obtained by following the vectors
$x_f\overrightarrow{a}$, $x_f\overrightarrow{b}$, and
$x_f\overrightarrow{c}$ in this order. We are going to show that all
these triangles $\Delta_f$ can be arranged as a tiling of $\Delta$,
forming a TC-representation of $T$ (i.e. such that $T= C(\R)$).

Note that a necessary condition for this to work is that (1) every
face of $f\in F_1$ around $a_0$, $b_0$, or $c_0$ is positive
(i.e. $x_f>0$), and that (2) for any inner vertex $v$ of $T$ its
positive (resp. negative) incident faces in $F_1$ are consecutive
around $v$. Otherwise, this would result in overlapping triangles
$\Delta_f$ (see Figure~\ref{fig:systemLE}). We first show that (1) and
(2) are fulfilled, and then we show that this suffices to ensure the
construction of $\R$.% (c.f. Claim~\ref{cl:drawing}).

Consider the incidence graph $I$, between vertices of $V(T)$ and faces
of $F_2$. First note that this plane graph has only size six faces and
that they are in bijection with the faces in $F_1$. Let us orient the
edges of $I$ as follows. An edge $vf$ of $I$, with $v\in V(T)$ and
$f\in F_2(T)$, is oriented from $v$ to $f$ if and only if the incident
faces (which correspond to faces in $F_1$) have different signs.  Note
that for an inner vertex of $T$, $d^+(v) = 2k$ for some $k\ge 1$ (as
$v$ is incident to positive and to negative faces in $T$), and that
$d^+(f)=1$ or $3$ for a face $f\in F_2$. The graph $I$ has $2n-2$
vertices (3 outer vertices of $T$, $n-3$ inner vertices of $T$, and
$n-2$ faces of $F_2$) and $3n-6$ edges. The outerface $f^o$ of $T$
has outdegree $3$ in $I$. Otherwise, among the three faces of $F_1$
incident to $a_0b_0$, $a_0c_0$, or $b_0c_0$ there would be positive
ones and negative ones. This would imply that two of $a_0$, $b_0$, and
$c_0$ have outdegree at least $2$ in $I$. This would be impossible as summing over the outdegrees of $a_0$, $b_0$, $c_0$, the inner vertices of $T$, and the faces of $F_2(T)$ would lead to
$2+2+ 2(n-3) + (n-2) > 3n-6$. We thus have that $d^+(f^o)=3$, and a
counting argument gives us that the other faces $f$ of $F_2$ have
outdegree one, that the outer vertices have outdegree zero, and that the 
inner vertices of $T$ have outdegree two. Thus, (1) and (2) are verified.

\begin{figure}%[h!]
\begin{center}
$\hspace{-0.5cm}$\includegraphics[scale=0.25]{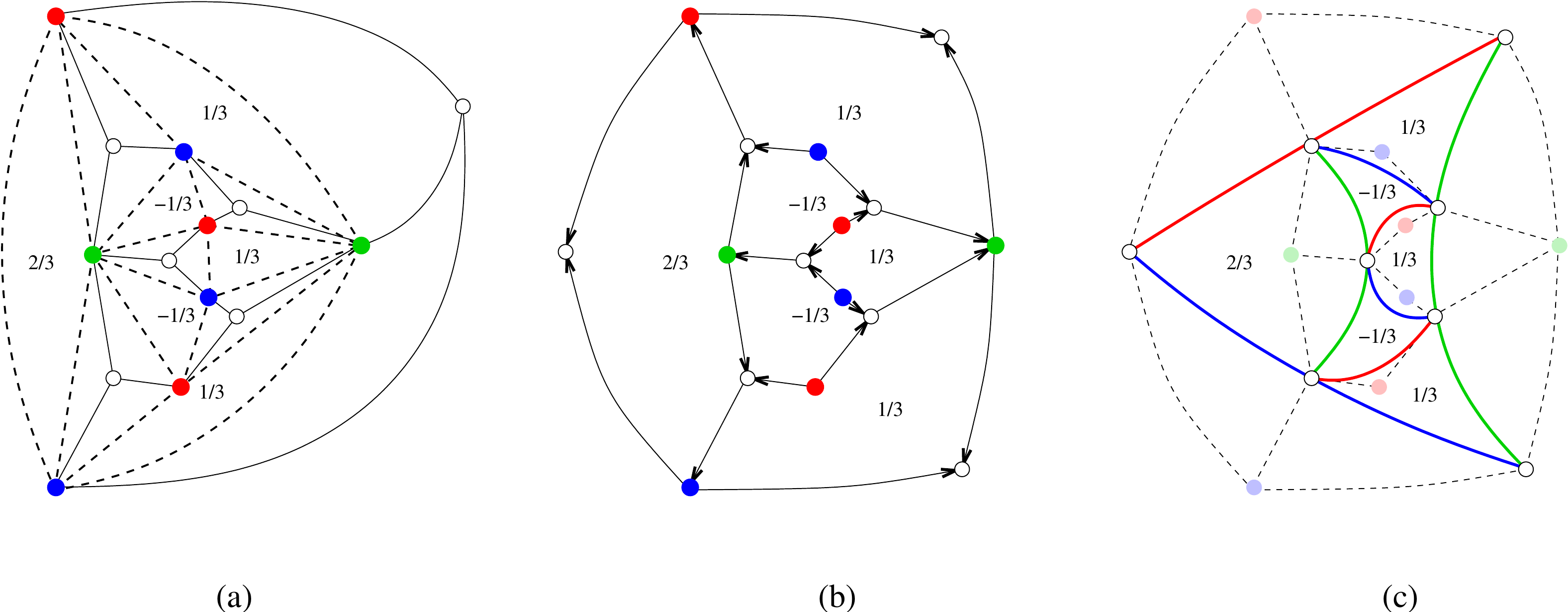}
\end{center}
\caption{(a) Example of a Eulerian triangulation $T$ (dashed lines),
  with incidence graph $I$. The numbers correspond to the solution of
  $\Ls$. (b) The graph $I'$ obtained after (Step 1). (C) The graph
  $G^\Delta$.}
\label{fig:II'}
\end{figure}

To construct the TC-representation of $T$, we define a plane graph
$G^\Delta$ from $I$ by replacing $f^o$ with three vertices (Step 1),
and for each vertex $v\in V(T)$, by turning its neighborhood in $I$
from a star into a path (Step 2).

\paragraph{(Step 1)}
The vertex $f^o$ is replaced by three new vertices $f^o_{\overline{a}}$,
$f^o_{\overline{b}}$, and $f^o_{\overline{c}}$ in such a way that
$f^o_{\overline{a}}$ is adjacent to $b_0$ and $c_0$ (see
Figure~\ref{fig:II'}). The six new edges are oriented towards the
newly created vertices.  Let us use $I'$ to denote this new oriented
graph. Note that now every vertex $v\in V(T)$ has outdegree two, and
that by assigning size $-1$ to the outerface, all faces incident to
$v$ sum up to zero.

\paragraph{(Step 2)}
For each vertex $v\in V(T)$, its neighborhood in $I'$ is turned into a
path $P_v$ whose ends are the out-neighbors of $v$. The in-neighbors
are ordered as follows in $P_v$. We first denote by $f^+$ (resp. $f^-$)
the out-neighbors of $v$ such that the face following $f^+$
(resp. $f^-$), around $v$ in clockwise order, has positive
(resp. negative) size ({\em i.e.} solution in $\Ls$). Two in-neighbors
$f,f'$ of $v$ are ordered along $P_v$ in such a way that $f$ is closer
to $f^+$ than $f'$, if and only if the sum of the face sizes going
around $v$ from $f^+$ to $f$ is lower than the sum from $f^+$ to
$f'$. If the two sums are equal, then $f$ and $f'$ are merged into a
single vertex (see Figure~\ref{fig:II'bis}). As all the faces around
$v$ have non-zero sizes, and as positive sizes are consecutive, a
vertex $f$ is merged at most once while constructing $P_v$. Note that $f$ is merged with another face $f'$ on $P_v$ only if $vf$ is oriented towards $v$ and this occurs for only one neighbor of $f$, so $f$ is not merged to more faces while constructing the other paths $P_u$.

The obtained plane graph is denoted
by $G^\Delta$.  Note that the inner faces of $G^\Delta$ correspond to
faces of $F_1$, and we assign them the corresponding sizes.  Note also
that a face of $G^\Delta$ corresponding to the face $a_ib_jc_k\in
F_1$, is bordered by three subpaths of paths $P_{a_i}$, $P_{b_j}$, and
$P_{c_k}$. We now assign a positive length to each edge of $G^\Delta$ so
that the length of each of these subpaths corresponds to the size of the face,
forgetting the sign. For each edge $ff'$ of a path $P_v$, we assign the
absolute value of the sum of the face sizes between $f$ and $f'$
around $v$ in $I'$.
\begin{figure}%[h!]
\begin{center}
%% $\hspace{-2cm}$
\includegraphics[scale=0.35]{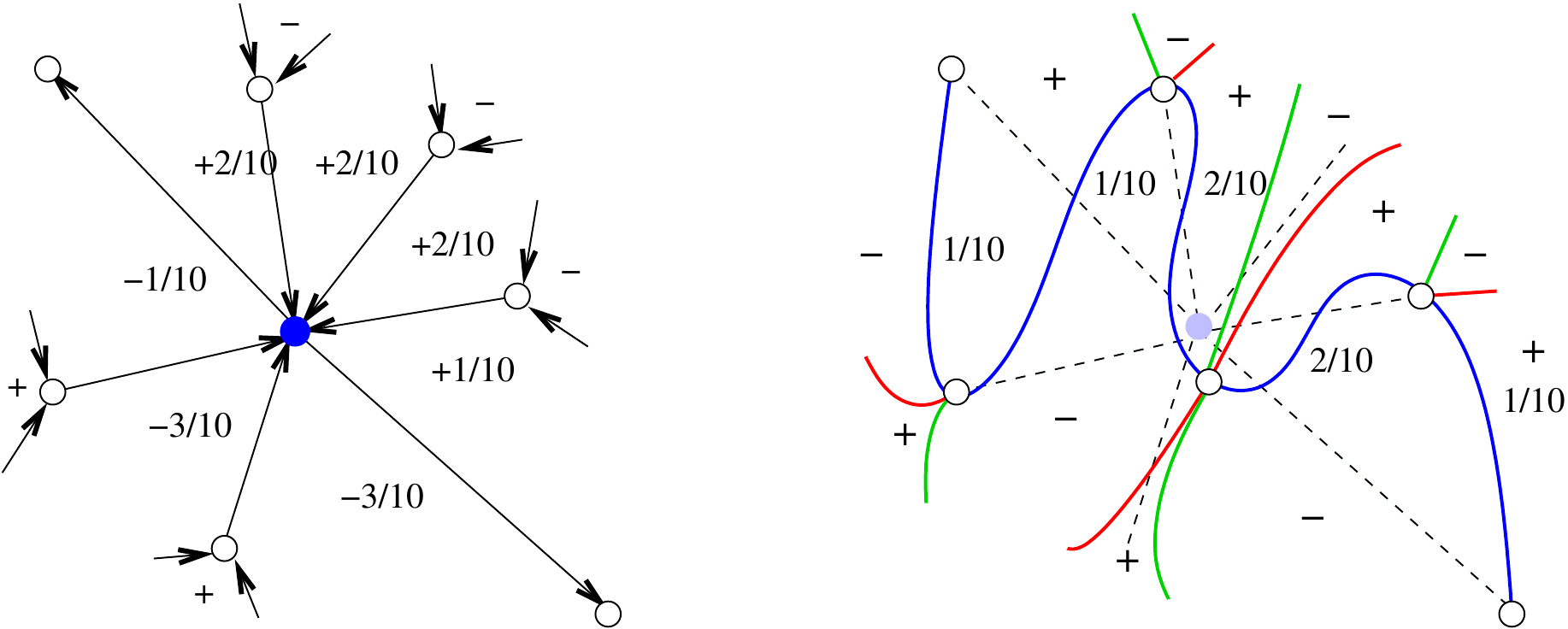}
\end{center}
\caption{An example of (Step 2) with a merge of $f$ and $f'$.}
\label{fig:II'bis}
\end{figure}

By construction, $G^\Delta$ has three types of vertices:
\begin{itemize}
  \item The vertices $f^o_{\overline{a}}$, $f^o_{\overline{b}}$, and
    $f^o_{\overline{c}}$, which have degree two. Indeed, {\em e.g.} the
    vertex $f^o_{\overline{a}}$ is at the end of $P_{b_0}$ and $P_{c_0}$.
    \item The vertices originating from a single vertex $f\in
      V(I)\setminus (V(T)\cup\{f^o\})$. As each such $f$ has in-degree two
      and out-degree one in $I$, it is at the end of two paths and in
      the middle of a third one.
    \item The vertices originating from two vertices $f,f'\in
      V(I)\setminus (V(T)\cup\{f^o\})$. By construction, such vertex
      is in the middle of a path, and has two paths ending on each
      side (corresponding to in-neighbors in $I$).
\end{itemize}
From the orientation of $I'$, note that the sign of the faces
alternate around any of these vertices (see Figure~\ref{fig:II'bis}).
We now want to draw $G^\Delta$ planarly, in such a way that its inner
faces are all homothetic to the triangle formed by following the three
vectors $\overrightarrow{a}$, $\overrightarrow{b}$, and
$\overrightarrow{c}$. More precisely, for a face $f$ of size $\alpha$
that is bordered by subpaths $P^f_a \subseteq P_{a_i}$, $P^f_b
\subseteq P_{b_j}$, and $P^f_c \subseteq P_{c_k}$, the subpaths
$P^f_a$, $P^f_b$ and $P^f_c$ should be mapped to vectors
$\alpha\overrightarrow{a}$, $\alpha\overrightarrow{b}$, and
$\alpha\overrightarrow{c}$, respectively, in such a way that the edge
length along these paths are met. Note that there is no local
obstruction to the existence of such embedding.
\begin{itemize}
  \item Each edge $ff'$ of $G^\Delta$ is consistently
    embedded. Indeed, the length of $ff'$ is set in $G^\Delta$, and
    regardless of the considered incident face (as these faces have
    different signs), the vector $\overrightarrow{ff'}$ has the same
    direction.
\item For the outer vertices $f^o_{\overline{a}}$,
  $f^o_{\overline{b}}$, and $f^o_{\overline{c}}$, their incident inner
  faces form an angle smaller than $\pi$ ({\em e.g.} for
  $f^o_{\overline{a}}$ the angle is the one from $\overrightarrow{c}$
  to $-\overrightarrow{b}$). For any other outer vertex $f$, which
  necessarily corresponds to a single vertex of $I'$, its (three)
  incident inner faces form an angle of size exactly $\pi$.
  For
  example, if $f$ is in the middle of the path $P_{a_0}$ and at the
  end of paths $P_{b_j}$ and $P_{c_k}$, we know by (1) that the inner
  faces incident to $P_{a_0}$ are positive, while the third one is
  negative because the edges $fb_j$ and $fc_k$ are oriented towards $f$
  in $I'$.  Thus, the angles around $f$ go from $\overrightarrow{a}$,
  to $-\overrightarrow{c}$, to $\overrightarrow{b}$, and to
  $-\overrightarrow{a}$.
\item For any inner vertex $f$ corresponding to a single vertex of
  $I'$, its (four) incident faces form an angle of size exactly
  $2\pi$.  For example, if $f$ is in the middle of a path $P_{a_i}$
  and at the end of paths $P_{b_j}$ and $P_{c_k}$, as the signs of the 
  four faces alternate, the angles around $f$ go from
  $-x\overrightarrow{a}$ to $x\overrightarrow{a}$, to
  $-x\overrightarrow{c}$, to $x\overrightarrow{b}$, and back to
  $-x\overrightarrow{a}$, for $x\in\{-1,+1\}$.
\item Similarly, for an inner vertex $f$ originating from two vertices
  of $I'$, the sum of the 6 angles is again $2\pi$.
\end{itemize}
%% \begin{figure}%[h!]
%% \begin{center}
%% %% $\hspace{-2cm}$
%% %\includegraphics[scale=0.35]{../fig/IIDeltaBis.pdf}
%% %\scalebox{0.6}{\input{../fig/I'GDelta.pdf_t}}
%% \end{center}
%% \caption{An example of (Step 2) with a merge of $f$ and $f'$.}
%% \label{fig:I'GDelta}%{fig:II'bis}
%% \end{figure} 
From these observations, a simple variant of Lemma 6 of~\cite{FSS18}
ensures the existence of such embedding. Alternatively, one could
triangulate $G^\Delta$ to use this lemma directly.

Note that this embedding is such that for each vertex $a_i\in V(T)$
(resp. $b_j\in V(T)$ and $c_k\in V(T)$) the corresponding path
$P_{a_i}$ (resp. $P_{b_j}$ and $P_{c_k}$) forms a segment parallel to
$\overrightarrow{a}$ (resp.  $\overrightarrow{b}$ and
$\overrightarrow{c}$). As in $G^\Delta$ a vertex $f$ is in the middle
of at most one path $P_v$, these segments do not cross. For
any inner edge of $T$, say $a_ib_j$ incident to a face $f\in F_2(T)$,
the paths $P_{a_i}$ and $P_{b_j}$ touch at the vertex $f$ of
$G^\Delta$. For the outer edges the contact points are
$f^o_{\overline{a}}$, $f^o_{\overline{b}}$, and
$f^o_{\overline{c}}$. We thus have a TC-scheme of $T$.
  
\paragraph{If some faces $f\in F_1$ are such that $x_f= 0$,} consider a face $a_ib_jc_k
\in F_1$ such that $x_{a_ib_jc_k}= 0$. Let $a',b',c'$ be the vertices belonging to a triangle adjacent to $a_ib_jc_k$ (see the left of Figure~\ref{fig:contraction}). Note that the vertices $a_i$ and $a'$ (resp. the vertices $b_j$ and $b'$, and the vertices $c_k$ and $c'$) have at least two common neighbors, $b_j,c_k$ (resp. $a_i,c_k$ and $a_i,b_j$). It may be the case that they have more common neighbors, but the planarity and the 3-coloring of $T$ forces one of these pairs to restrict to these two common neighbors only. Let us assume that $a_i$ and $a'$ have exactly two common neighbors, $b_j,c_k$. Let $T'$ be the Eulerian triangulation obtained from $T$ by
deleting the edges $b_jc_k$, $a'b_j$ and $a'c_k$, and by
merging $a_i$ and $a'$ (see Figure~\ref{fig:contraction}). The resulting vertex of $T'$ is 
denoted by $a'_i$. Note that by the choice of $a_i,a'$, $T'$ has no multiple edges. We can thus apply the induction on $T'$.

\begin{figure}
    \centering
    \includegraphics[width=0.75\linewidth]{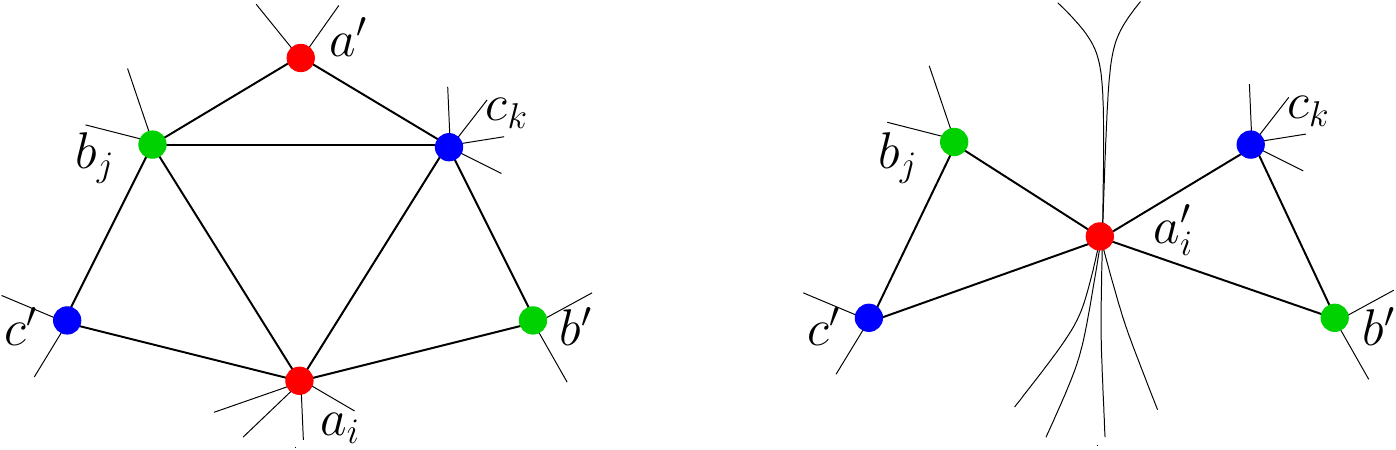}
    \caption{Reduction to a smaller triangulation $T'$ by identifying $a_i$ and $a'$.}
    \label{fig:contraction}
\end{figure}
Let $\Ls'$ be the linear system defined for $T'$. Note
that a solution of $\Ls$ clearly induces a solution of $\Ls'$. Indeed,
every vertex $v\in V(T')\setminus\{a'_i,b_j,c_k\}$ is incident to the
same faces as in $T'$, so they sum up to $0$ (or to $1$ for
outer vertices). For $b_j$, or $c_k$ these vertices are incident to
one less face of $F_1$, the face $a_ib_jc_k$, and as $x_{a_ib_jc_k}=
0$, their incident faces still sum up to $0$ (or to $1$) in
$T'$. Similarly, as the faces of $F_1$ incident to $a'_i$ in $T'$ are
the faces of $F_1$ incident to $a_i$ or to $a'$ in $T$, except
$a_ib_jc_k$, they sum up to $0$. As the solution of $\Ls'$ has one
less $0$ entry, we can apply the induction, and consider a TC-scheme
$\R'$ of $T'$ corresponding to this solution of $\Ls'$. We consider
different cases according to whether $a_i$ and $a'$ have non-zero
valued incident faces in $T$.

\paragraph{If $a'$ has only zero valued incident faces in $T$,}
then $\R'$ is also a TC-scheme for $T$ corresponding to $\Ls$. Note by Lemma~\ref{lem:equiv-def-scheme}.(2) that every non-degenerate inner face $f$ of $C(\R')$ is a face of $F_1(T')$. Renaming $a_i$ the vertex originated by the merging of $a_i$ and $a'$, instead of $a'_i$, all the faces of $F_1(T')$ with non-zero size, are also faces of $F_1(T)$. Hence, $\R'$ is also a TC-scheme for $T$. The fact that it corresponds to $\Ls$ is clear by the construction.

\paragraph{If $a_i$ has only zero valued incident faces in $T$,} we proceed as above. The only difference is that we have to denote $a'$ the 
vertex originated by the merging of $a_i$ and $a'$.

\paragraph{If both $a_i$ and $a'$ have non-zero valued incident faces in $T$,}
let us divide the segment ${\bf a'_i}$ of $\R'$ into two parts, one
for each of $a_i$ and $a'$. Note that the faces of $F_1(T)\setminus
\{a_ib_jc_k\}$ incident to $a_i$ (resp. $a'$) in $T$ correspond to
consecutive triangles around ${\bf a'_i}$. Furthermore, as their sizes
sum up to $0$ there is a point ${\bf p} \in {\bf a'_i}$ that divides
${\bf a'_i}$ into two parts, ${\bf a_i}$ and ${\bf a'}$, such that
the faces of $F_1\setminus \{a_ib_jc_k\}$ incident to $a_i$
(resp. $a'$) in $T$ correspond to triangles with a side contained
inside ${\bf a_i}$ (resp. ${\bf a'}$). Let us denote by $\R$ the
obtained TC-representation. As every non-degenerate face $f$ of $C(\R)$
corresponds to a face of $F_1(T)$ whose size is $x_f$, by
Lemma~\ref{lem:equiv-def-scheme}.(2) we have that $\R$ is a TC-scheme of $T$
following the solution of $\Ls$.  This concludes the induction step of
the proof.
\end{proof}

\section{3-slopes segment representations}\label{sec:3-dir}

In this section, we use Theorem~\ref{thm:TC} to prove the main theorem
of the article, Theorem~\ref{thm:main}. As already mentioned, it is
sufficient to prove it for Eulerian triangulations. From now on, given a Eulerian triangulation with more than three vertices, let $a_1$,
$b_1$ and $c_1$ be the vertices forming an inner face with vertices
$b_0$ and $c_0$, with $a_0$ and $c_0$, and with $a_0$ and $b_0$,
respectively. 
Theorem~\ref{thm:main} follows from the following technical
proposition.

%% Given a
%% TC-scheme $\R$ of $T$, note that $a_1$ is not included in a degenerate
%% face of $C(\R)$. Indeed, such a face should involve $b_0$ and $c_0$
%% but their segments intersect at a point where no other segment
%% intersects.  This also holds for $b_1$ and $c_1$, and those vertices
%% are thus part of $C(\R)$. We can hence

\begin{figure}%[h!]
\begin{center}
\includegraphics[scale=0.25]{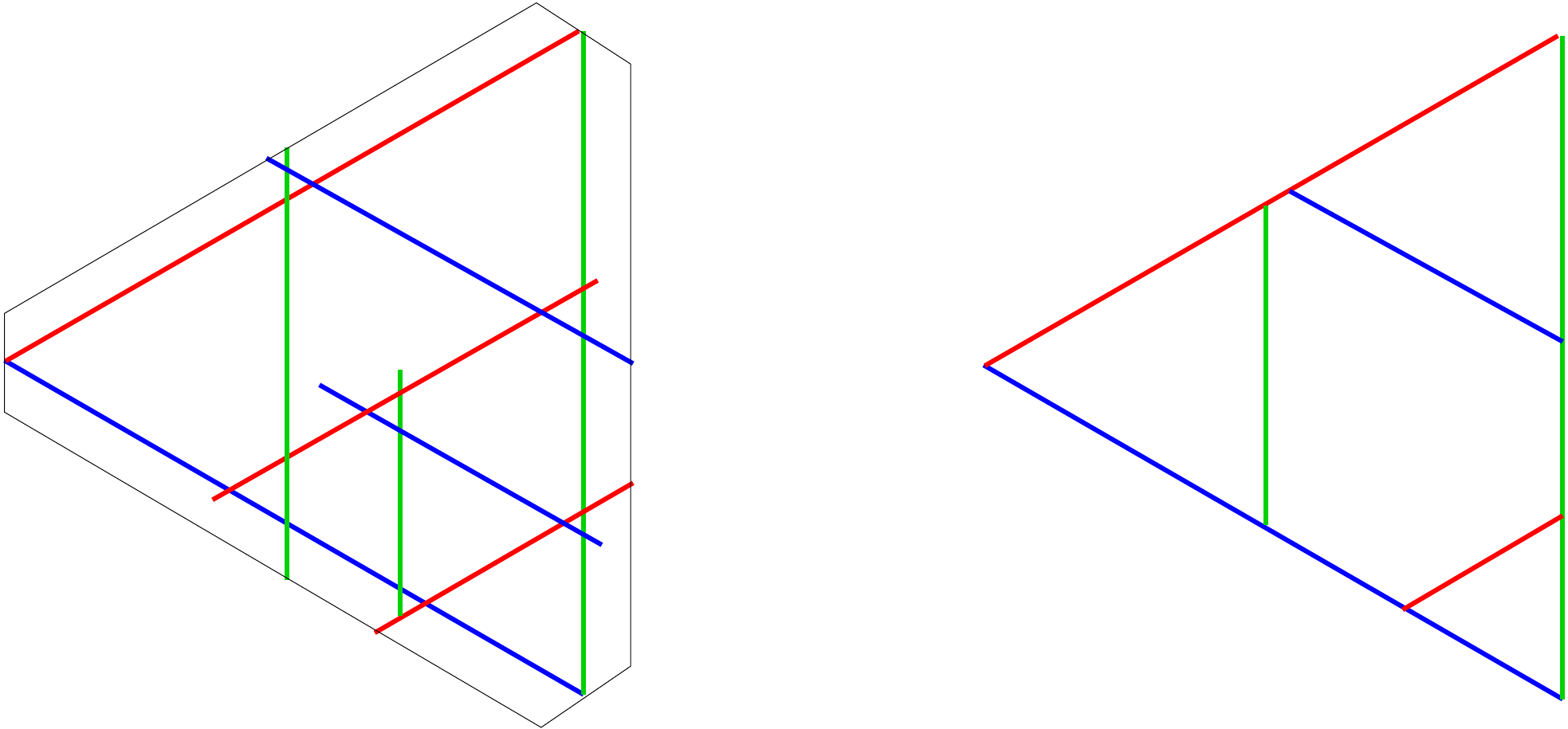}
\end{center}
\caption{(left) A $3$-slopes segment representation inside a
  hexagon. Note that some intersection points lie inside the triangle ${\bf a_0b_0c_0}$, while some are outside this triangle (here this is the case for the intersection of ${\bf b_1}$ and ${\bf c_1}$). (right) An illustration of its {\em shape}.}
\label{fig:prop}
\end{figure}

\begin{proposition}\label{prop:technical}
For every sufficiently small $\epsilon>0$, every simple Eulerian triangulation $T$
admits a 3-slopes segment representations $\R$ such that:
\begin{itemize}
\item The segments ${\bf a_0}$, ${\bf b_0}$, and ${\bf c_0}$ form a triangle $\Delta$ of size $1$ (its sides are obtained by following $\overrightarrow{a}$, $\overrightarrow{b}$,
and $\overrightarrow{c}$).
\item Every segment is contained in the hexagon centered on $\Delta$, obtained by successively following $(1-\epsilon)\overrightarrow{a}$, $-2\epsilon\overrightarrow{c}$, $(1-\epsilon)\overrightarrow{b}$, $-2\epsilon\overrightarrow{a}$, $(1-\epsilon)\overrightarrow{c}$, and $-2\epsilon\overrightarrow{b}$ (see Figure~\ref{fig:prop}).
\item The segments ${\bf a_1}$, ${\bf b_1}$, and
${\bf c_1}$ have both endpoints on the border of the hexagon.
\item No three segments intersect at the same point. 
\item For every inner edge $uv$ the segments ${\bf u}$ and ${\bf v}$ cross each other (i.e. their intersection point is not an endpoint).
\end{itemize}
\end{proposition}

Given such representation $\R$ of a triangulation having some inner
vertices, we define the \emph{shape} of $\R$ as the triplet
$(s_a,s_b,s_c)$ of sizes  of the triangles corresponding to
$a_1b_0c_0$, $a_0b_1c_0$, $a_0b_0c_1$, respectively, in $\R$  (see the right part of Figure~\ref{fig:prop}).\\

\begin{proof}
We proceed by induction on $|V(T)|$. The initial case
of this induction, when $|V(T)|=3$ clearly holds.
We thus proceed to the induction step with a Eulerian triangulation $T$ such that $|V(T)|>3$, and we assume that the proposition holds for
any simple Eulerian triangulation with fewer vertices. 

By Theorem~\ref{thm:TC}, we consider a TC-scheme $\R$ of $T$, and
we prolong the segments ${\bf a_1}$, ${\bf b_1}$, and
${\bf c_1}$ to have their endpoints on the border of the hexagon.
Then, to reach the sought representation, we {\em resolve} every {\em degenerate point} of $\R$. A {\em $k$-degenerate point} of $\R$, for $k=3,5$ or 6, is a point ${\bf p}$ belonging to $k$ segments.  Here, {\em resolving} means that the
segments of a 3-degenerate point (resp. a 5- or a 6-degenerate point) are
moved to form a triangle (resp. two triangles, or a polygon) inside which we are going
to draw a 3-slopes representation of the graph corresponding to this
degenerate face of $C(\R)$, this is possible by using the induction on
this smaller graph. The degenerate points of $\R$ are resolved from
left to right. This means that at a given stage of this process there
is a vertical line (parallel with $\overrightarrow{b}$) $\V$ such that on its
left all the intersection points (except for the one of $a_0c_0$) are crossing points. This
implies that on the left of $\V$ the representation handles some small
perturbations: Every segment, except ${\bf a_0}$ and ${\bf c_0}$, can slightly move in any direction, without creating or destroying any intersection point.

Let $\V$ be the leftmost vertical line containing degenerate
points. We resolve those degenerate points by slightly moving segments
on the left of or on $\V$, while maintaining the right side of the
representation unchanged (except at the vicinity of $\V$). We consider different cases according to the
degenerate points on $\V$. 

\begin{figure}%[h!]
\begin{center}
\includegraphics[scale=0.35]{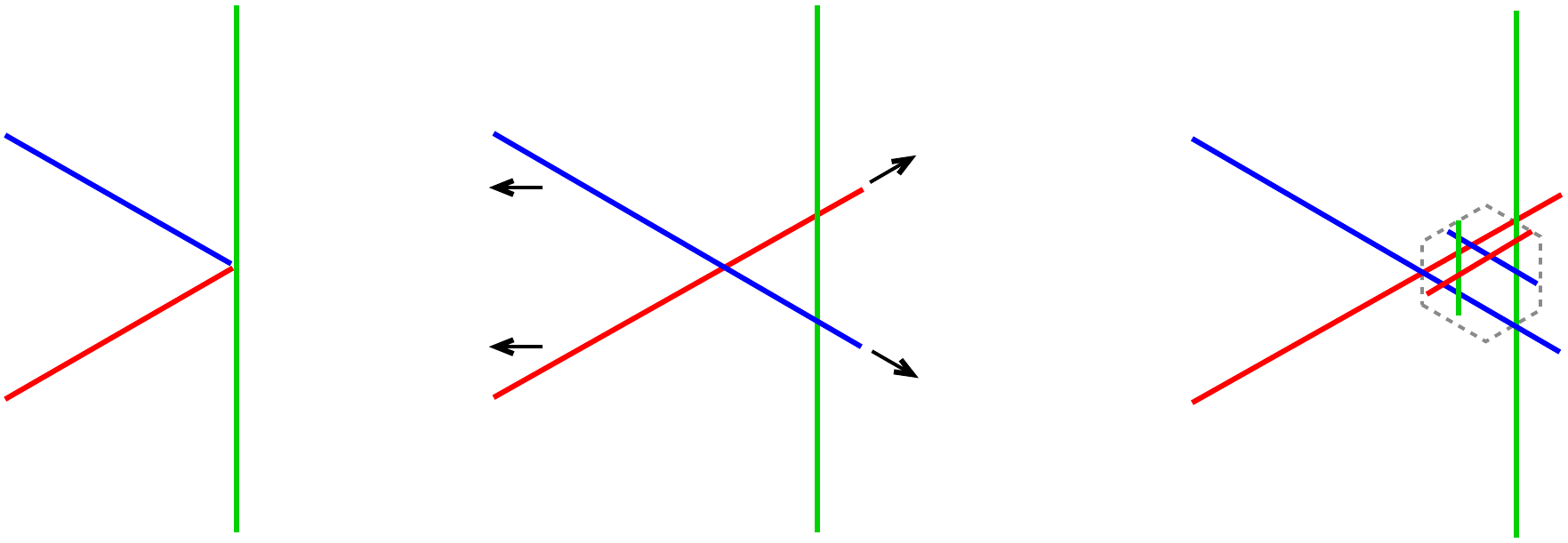}
\end{center}
\caption{(left) A 3-degenerate point on $\V$ (middle) Small perturbation of $\R$ (right) The addition of a representation roughly inside the new triangle. The new segments are added in the hexagon with dashed gray border.}
\label{fig:resolve-3pt-left}
\end{figure}

\paragraph{If $\V$ contains a 3-degenerate point ${\bf p}$ in the interior of a
(vertical) segment ${\bf b_j}$ and at the end of two segments ${\bf
    a_i}$ and ${\bf c_k}$ lying on the left of $\V$,} the situation is
    rather simple. Move these segments ${\bf
    a_i}$, ${\bf c_k}$ a little to the left and
    slightly prolong them to intersect ${\bf b_j}$ (see
    Figure~\ref{fig:resolve-3pt-left}). As there is no degenerate
    point on the left of $\V$ these moves can be done while
    maintaining the existing intersections and avoiding new
    intersections. If $a_ib_jc_k$ is not a face of $T$, consider the
    triangulation $T'$ induced by the vertices in the cycle
    $a_ib_jc_k$ of $T$. By induction, $T'$ has a representation that
    can be drawn inside a hexagon centered on the newly formed triangle bordered by the
    segments ${\bf a_i}$, ${\bf b_j}$ and ${\bf c_k}$.

\begin{figure}%[h!]
\begin{center}
\includegraphics[scale=0.35]{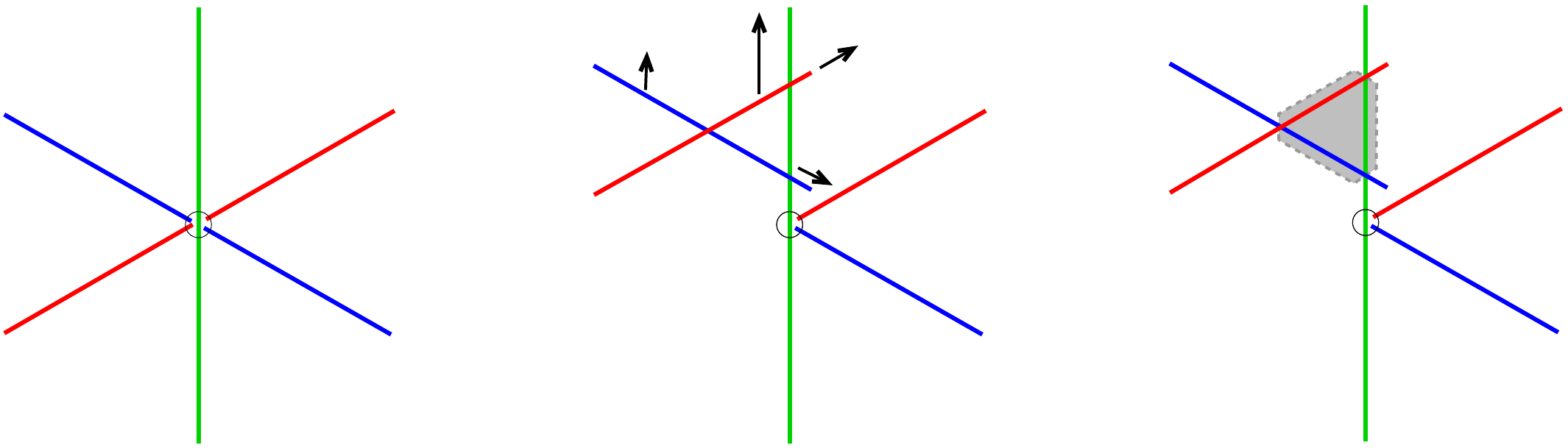}
%resolve-3pt-double-3.pdf}
\end{center}
\caption{(left) A 5-degenerate point on $\V$.  (middle) A small perturbation of $\R$. The small circle indicates the position of the original 5-degenerate point ${\bf p}$, that is now a simple 3-degenerate point. (right) Addition of a 3-slope segment representation, if needed.}
\label{fig:resolve-3pt-double}
\end{figure}

\paragraph{If $\V$ contains a 5-degenerate point ${\bf p}$ in the interior of a
(vertical) segment ${\bf b_j}$,} the situation is similar to the
    previous one. Move the segments on the left of $\V$ as depicted in
    Figure~\ref{fig:resolve-3pt-double}. If the new triangle is not a
    face of $T$, we add a representation inside. We are now left with
    a single 3-degenerate point at ${\bf p}$. This corresponds to the
    following case.

\begin{figure}%[h!]
\begin{center}
\includegraphics[page=2,scale=0.35]{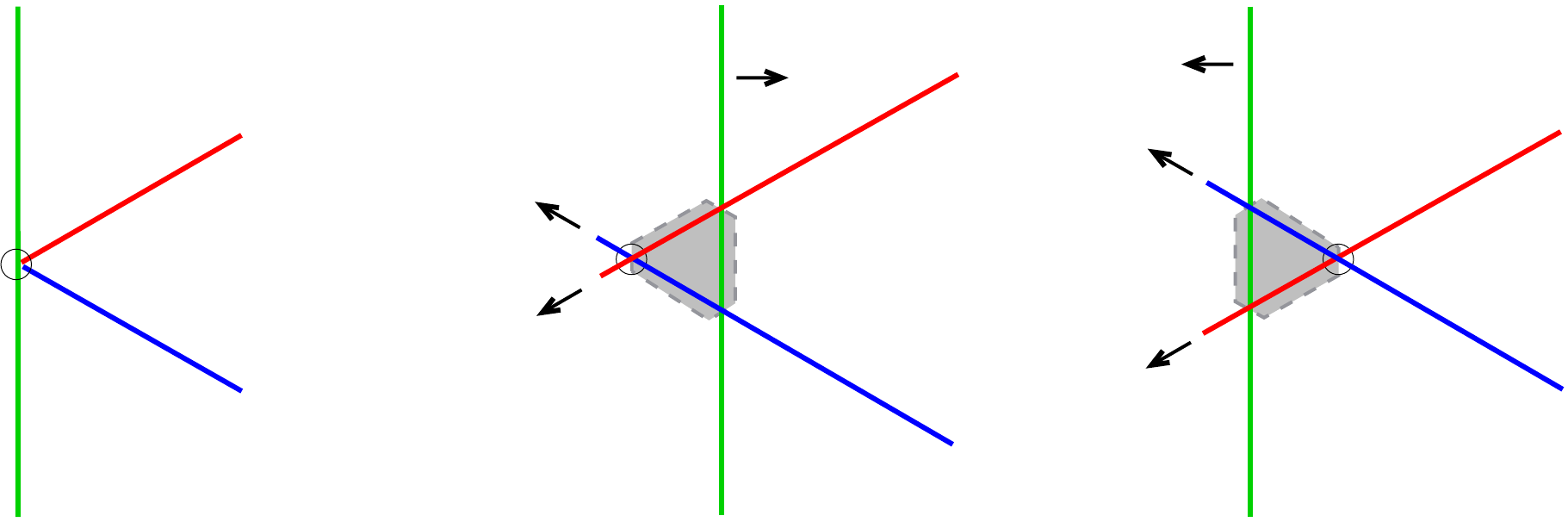}
\end{center}
\caption{(left) A 3-degenerate point ${\bf p}$, in a small circle, with segments on the right of $\V$. (middle) Slightly moving ${\bf b_j}$ to the right, and addition of extra segments in the hexagon, if needed. (right) Slightly moving ${\bf b_j}$ to the left.}
\label{fig:resolve-3pt-right}
\end{figure}

\paragraph{If $\V$ contains a 3-degenerate point ${\bf p}$ in the interior of a
(vertical) segment ${\bf b_j}$ and at the end of two segments, ${\bf a_i}$ and ${\bf c_k}$, lying on the right of $\V$,} one can move
    ${\bf b_j}$ slightly to the right or slightly to the left and
    resolve these points without changing the right part of the
    representation. The choice of moving ${\bf b_j}$ to the right or
    to the left is explained in the next paragraph, but we can
    assume this move to be arbitrarily small. Regardless of the direction
    ${\bf b_j}$ is moved, one has to prolong ${\bf a_i}$ and ${\bf
    c_k}$ to keep the intersections between these segments and ${\bf b_j}$ (see Figure~\ref{fig:resolve-3pt-right}). Note
    that in order to preserve the representation on the right of $\V$
    the segments ${\bf a_i}$ and ${\bf c_k}$ are not moved, they are
    only prolonged across ${\bf p}$. Again, if $a_ib_jc_k$ is not a
    face of $T$, we draw a representation inside the newly
    formed triangle. Note that if ${\bf b_j}$ moves to the right, the
    triangle bordered by ${\bf a_i}$, ${\bf b_j}$ and ${\bf c_k}$ has
    negative size, but it suffices to apply a homothety with negative
    ratio to obtain a representation that can be drawn inside.

\paragraph{Consider now the degenerate points at the end of a (vertical) segment
${\bf b_j}$ of $\V$.} Let ${\bf b_1},{\bf b_2},\ldots,{\bf b_t}$ be a
maximal sequence of segments on $\V$, from bottom to top, such that ${\bf b_j}$ and ${\bf
b_{j+1}}$ intersect in a point. We are going to move these segments slightly to the right or slightly to the left of $\V$. The direction of these moves, right or left, and their exact magnitude will be set later, but remember that these moves can be made arbitrarily small. Recall that the 3-degenerate points in the interior of the
segments ${\bf b_j}$ with $1\le j\le t$ can be dealt if the move of
${\bf b_j}$ is sufficiently small (see the previous cases). 

\begin{figure}
    \centering
    \includegraphics[width=0.9\linewidth]{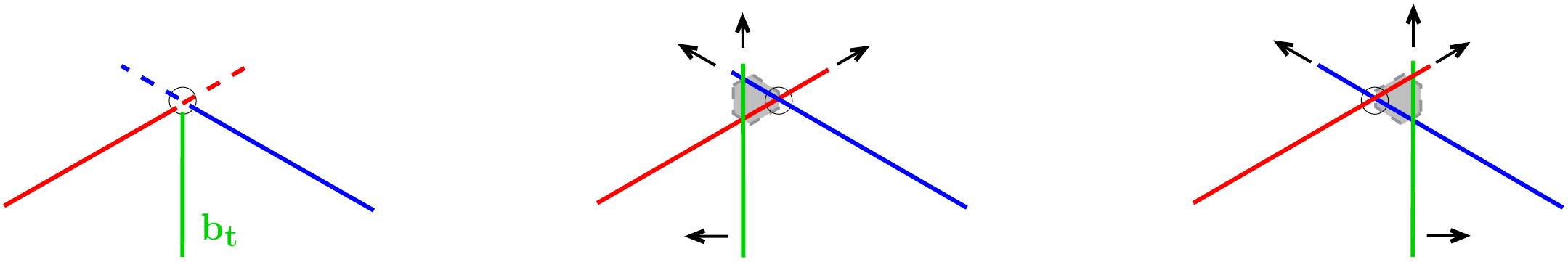}
    \caption{(left) A 3-degenerate point ${\bf p}$ at the end of ${\bf b_t}$. The cases where ${\bf a_i}$ end at ${\bf p}$ is similar to the case where ${\bf c_k}$ ends at this point, since the segment ending there is prolonged. (middle) Slightly moving ${\bf b_t}$ to the left, and addition of extra segments in the hexagon, if necessary. (right) Slightly moving ${\bf b_t}$ to the right.}
    \label{fig:resolve-bt}
\end{figure}
Consider first the upmost endpoint ${\bf p}$ of ${\bf b_t}$, and let ${\bf a_i}$ and ${\bf c_k}$ be the other segments meeting at this point. The point ${\bf p}$ is an endpoint for one of ${\bf a_i}$ or ${\bf c_k}$. We slightly prolong this segment across ${\bf p}$ (see Figure~\ref{fig:resolve-bt}). Then, whatever the direction ${\bf b_t}$ is moved, by
slightly prolonging ${\bf b_t}$, we maintain the desired intersection. Note that these prolongings ensure that the representation on the right of $\V$ is unchanged, such as segment ${\bf a_0}$ if ${\bf a_i} = {\bf a_0}$. Again, if $a_ib_tc_k$ is not a face of $T$, we draw a representation inside the newly formed triangle. In this case, if ${\bf a_i} = {\bf a_0}$, note that as the move of ${\bf b_t}$ can be arbitrarily small, the representation of $T$ can remain inside the hexagon of Proposition~\ref{prop:technical}.
The case of the lowest endpoint of ${\bf b_1}$ is symmetrical.

Now we arbitrarily choose to move ${\bf b_1}$ towards the left of $\V$, and we assume that move to be arbitrarily small. In the following we will see how the direction chosen for this move, to the left, implies a direction for moving ${\bf b_2}$, which in turn implies a direction for moving ${\bf b_3}$, and so on until ${\bf b_t}$. Regarding the magnitude of these moves, they are all positively correlated, meaning that all these moves can be arbitrarily small, if the first one (of ${\bf b_1}$) is chosen sufficiently small.

Consider now any
intersection point ${\bf p}$ between ${\bf b_j}$ and ${\bf
b_{j+1}}$, and let us distinguish various cases according to the other segments containing ${\bf p}$. 

\begin{figure}%[h!]
\begin{center}
\includegraphics[scale=0.32]{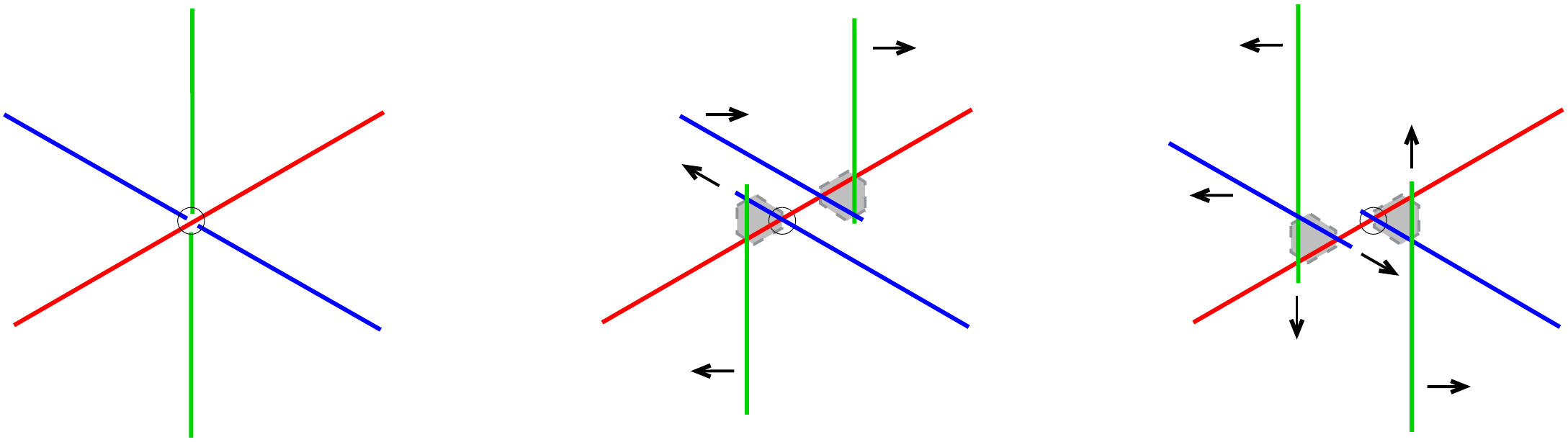}
  \end{center}
\caption{(left) A 5-degenerate point on $\V$ (middle) \& (right) Small moves that resolve this point with respect to the directions taken by ${\bf b_j}$ and ${\bf b_{j+1}}$.}
\label{fig:resolve-3+3pt}
\end{figure}

Suppose there is a segment ${\bf a_i}$ going through ${\bf p}$. It is shown
in Figure~\ref{fig:resolve-3+3pt} how to resolve these two overlapped
3-degenerate points, in order to create two triangles, where one can
add a small representation if needed.  The case where there is a
segment ${\bf c_k}$ going through ${\bf p}$ is similar.

% \begin{figure}%[h!]
% \begin{center}
% \scalebox{0.25}{\input{resolve-6pt.pdf_t}}
% \end{center}
% \caption{From left to right : A 6-degenerate point on $\V$. Resolution when there is no chord in $b_jacb_{j+1}a'c'$ with the shape of $\R'$. Resolution when none of $b_{j}c$, $ca'$, or $a'b_{j}$ is a chord,  with the shape of $\R'$. Resolution when $ac'$ and $ca'$ are chords, with the shape of $\R_2$.}
% \label{fig:resolve-6pt}
% \end{figure}
\begin{figure}%[h!]
\begin{center}
\includegraphics[scale=0.3]{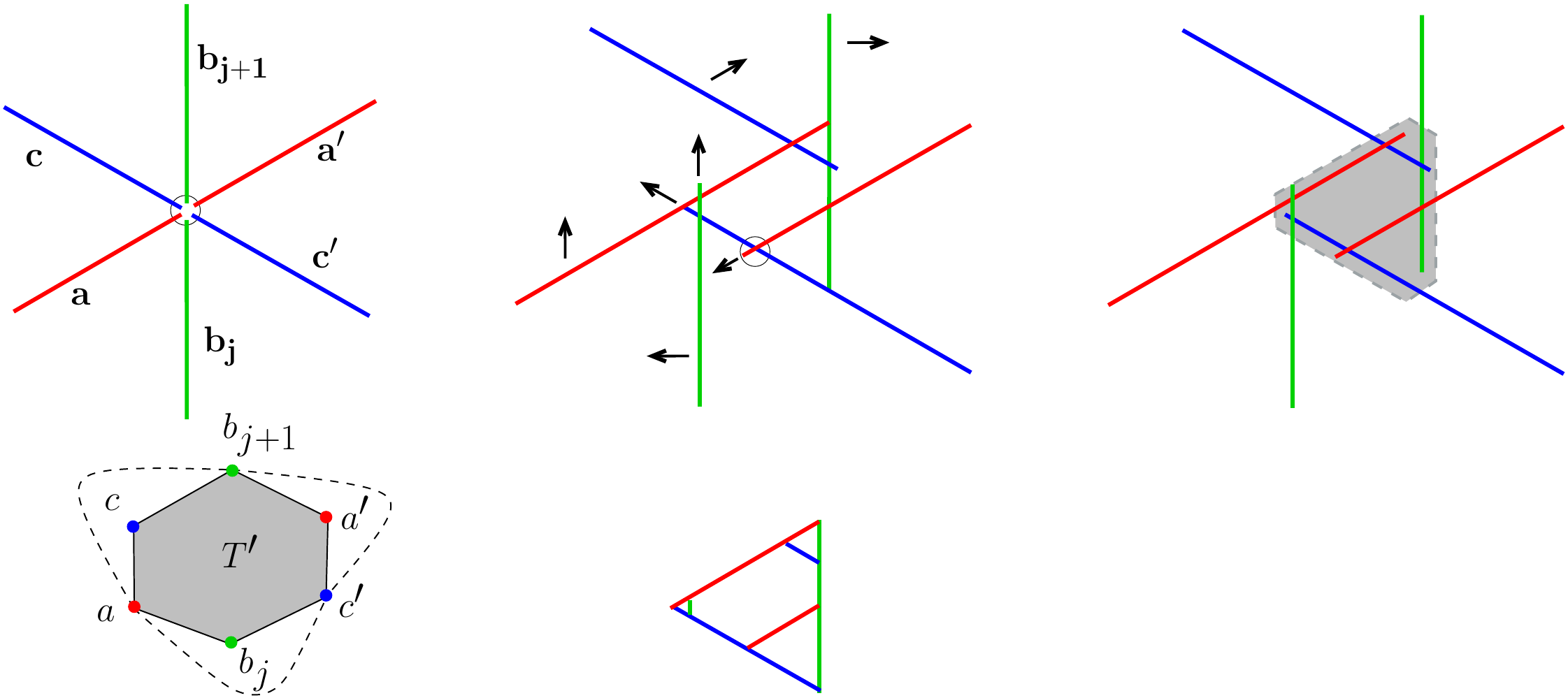}
%\scalebox{0.25}{\input{resolve-6pt.pdf_t}}
\end{center}
\caption{Resolution of a 6-degenerate point ${\bf p}$ when there is no chord $ab_{j+1}$, $b_{j+1}c'$, or $ac'$.
(left) A 6-degenerate point on $\V$, and below, the triangulation $T'$ obtained from the vertices inside this degenerate face of the TC-scheme. (middle) The shape of $\R'$  and the moves made according to this shape and according to the move of ${\bf b_j}$ towards the left. (right) The 3-slope segment representation $\R'$ is included, and the segments ${\bf a}$, ${\bf b_{j+1}}$, and ${\bf c'}$ are shortened to avoid the artificially added edges $ab_{j+1}$, $b_{j+1}c'$, and $ac'$.}
\label{fig:resolve-6pt-1-24}
\end{figure}

Assume now that six segments intersect at ${\bf p}$. Let ${\bf b_j}$
be the one below ${\bf p}$, and let ${\bf a}$, ${\bf c}$, ${\bf
  b_{j+1}}$, ${\bf a'}$, and ${\bf c'}$ be the other ones around ${\bf
  p}$ clockwisely. The
degenerate face corresponding to ${\bf p}$ in $C(\R)$ is bounded by the cycle $(b_j,a,c,b_{j+1},a',c')$, and there are several cases according to whether there are chords within this cycle in $T$.
For all the cases, we assume that the move of ${\bf b_j}$, its direction and its (arbitrarily small) magnitude is set, and we will see how it determines the direction and magnitude of the moves applied to ${\bf a}$, ${\bf c}$, ${\bf b_{j+1}}$, ${\bf a'}$, and ${\bf c'}$.

Suppose that none
of $ab_{j+1}$, $b_{j+1}c'$, or $ac'$ is a chord of $(b_j,a,c,b_{j+1},a',c')$ in $T$  (see Figure~\ref{fig:resolve-6pt-1-24}), then we consider
the subgraph of $T$ induced by the vertices on and inside this cycle.
We add the edges $ab_{j+1}$, $b_{j+1}c'$, and $ac'$ outside the
cycle, and we denote by $T'$ the obtained simple Eulerian
triangulation. By the induction, we know that $T'$ admits a 3-slope
segment representation $\R'$, and let $(s_a,s_b,s_c)$ be the shape of
$\R'$. If ${\bf b_{j}}$ is moved to the left of $\V$, we resolve the point ${\bf p}$ by moving the segments as depicted in
Figure~\ref{fig:resolve-6pt-1-24}. Each of these moves are proportional to the move of ${\bf b_j}$ towards the left, and the magnitude of each of these moves
is prescribed by the shape $(s_a,s_b,s_c)$ in order to allow us to
copy $\R'$ inside the triangle formed by ${\bf a}$, ${\bf b_{j+1}}$,
and ${\bf c'}$. Then we shorten ${\bf a}$, ${\bf b_{j+1}}$, and ${\bf
c'}$ to avoid intersections among them. The case where ${\bf b_{j}}$ is moved to the right of $\V$ is depicted in the left of Figure~\ref{fig:resolve-6pt-2-24}. In that case, note that $\R'$ is not  only rescaled to fit into the hexagon, but we first consider a homothety of $\R'$ with negative ratio. 

If none of $b_{j}c$, $ca'$, or $a'b_{j}$ is a chord of
$(b_j,a,c,b_{j+1},a',c')$ we proceed similarly (see the middle and the right of Figure~\ref{fig:resolve-6pt-2-24}). The only difference is that
we add the edge $b_{j}c$, $ca'$, or $a'b_{j}$ outside
$(b_j,a,c,b_{j+1},a',c')$ to obtain $T'$.

\begin{figure}
    \centering
    \includegraphics[width=\linewidth]{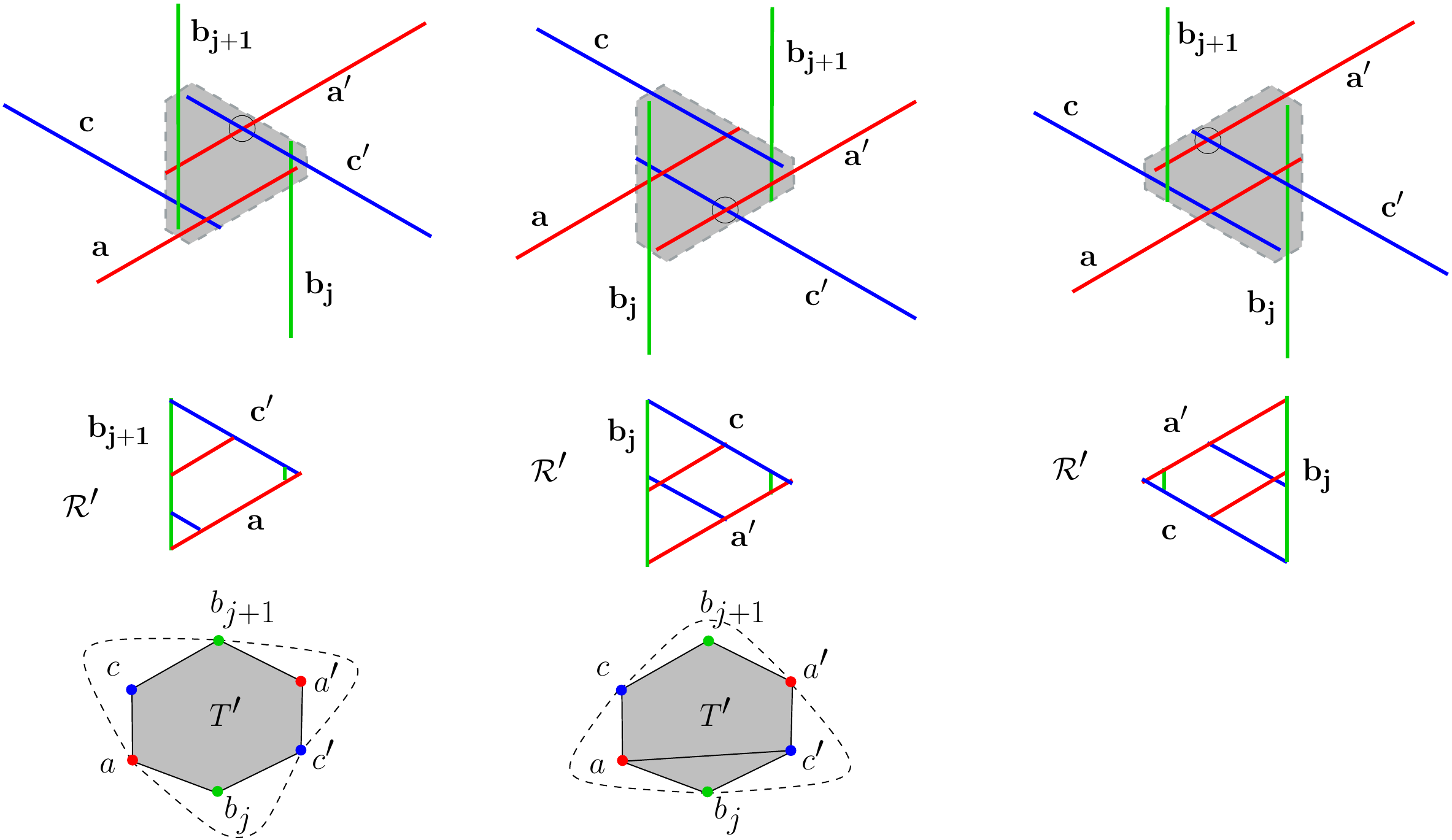}
    \caption{(left) Resolution of a 6-degenerate point ${\bf p}$ when there is no chord $ab_{j+1}$, $b_{j+1}c'$, or $ac'$, and when ${\bf b_j}$ is moved to the right.
    (middle) The case where none of $b_{j}c$, $ca'$, or $a'b_{j}$ is a chord and where ${\bf b_j}$ is moved to the left. In this illustration there is a chord $ac'$, but this may not be the case, of course. (right) Same case with ${\bf b_j}$ moved to the right of ${\bf p}$.}
    \label{fig:resolve-6pt-2-24}
\end{figure}

We now have to deal with the case where $(b_j,a,c,b_{j+1},a',c')$ has two opposite chords.
If $ac'$ and $a'c$ are chords (see the left part of Figure~\ref{fig:resolve-6pt-3-24}), we consider two triangulations. Let $T_1$ be the one
inside the cycle $(c,b_{j+1},a')$ and let $T_2$ be the one obtained from the
interior of the $5$-cycle $(a',c',b_{j},a,c)$ by adding the edges
$a'b_{j}$ and $b_{j}c$ outside. By the induction, we know that $T_1$ and
$T_2$ admit 3-slopes segment representations $\R_1$, and $\R_2$.  We resolve the point ${\bf p}$ by
moving the segments as depicted in Figure~\ref{fig:resolve-6pt-3-24}, and
the magnitude of each of these moves, except for ${\bf b_j}$ and ${\bf b_{j+1}}$, is
prescribed by the shape of
$\R_2$. Then we shorten two segments to avoid
the intersections corresponding to $a'b_{j}$ and $b_{j}c$. The
segment ${\bf b_{j+1}}$ is moved away to avoid the
interior of the hexagon containing $\R_2$. Then $\R_1$ is drawn
inside the triangle bordered by ${\bf b_{j+1}}$, ${\bf a'}$ and ${\bf
c}$. This is possible because the hexagons containing  $\R_1$ and $\R_2$ do not overlap.

If $b_jc$ and $b_{j+1}c'$ are chords, we proceed similarly (see the right part of Figure~\ref{fig:resolve-6pt-3-24}). The difference is that, here,  $T_1$ is the triangulation 
inside the cycle $(b_{j+1},a',c')$ and let $T_2$ be the one obtained from the
interior of the $5$-cycle $(c',b_{j},a,c,b_{j+1})$
by adding the edges $ab_{j+1}$ and $ac'$ outside.

\begin{figure}
    \centering
    \includegraphics[width=\linewidth]{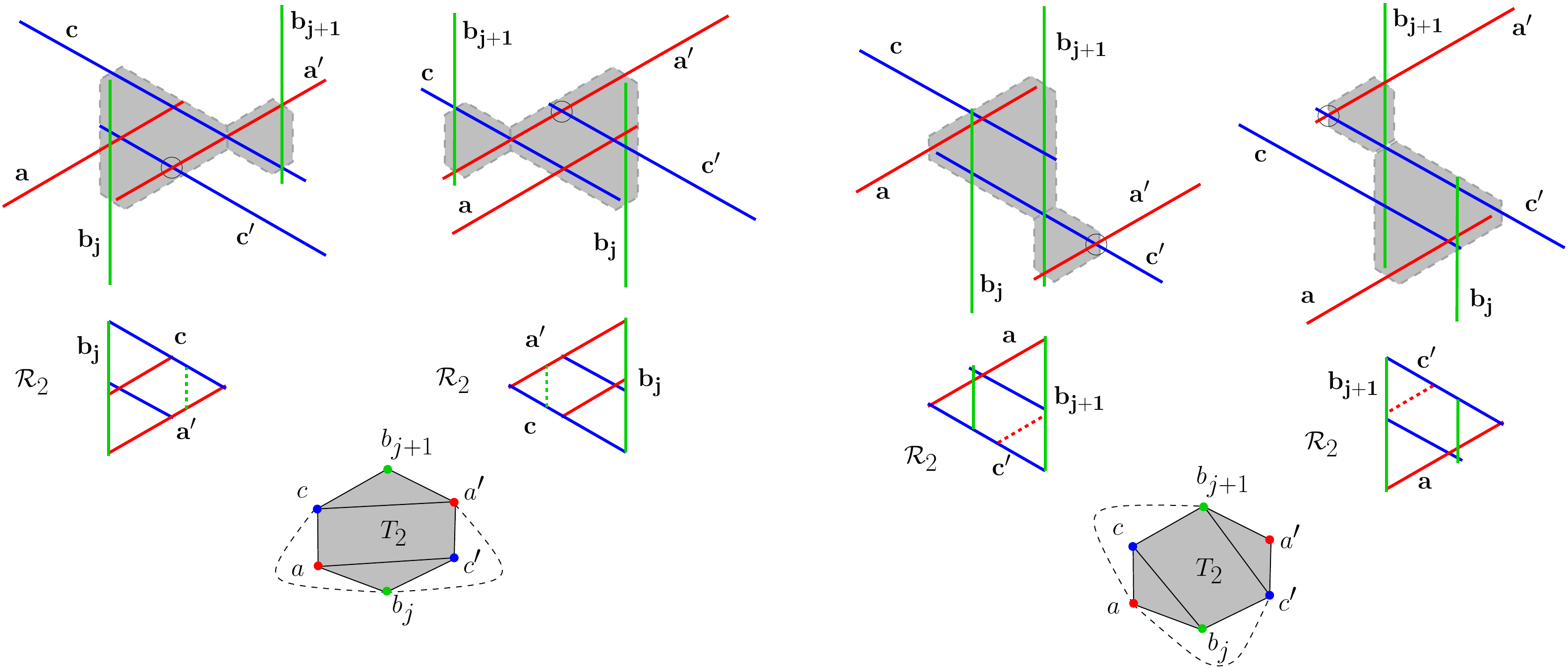}
    \caption{(left) The case where $ac'$ and $a'c$ are chords, with ${\bf b_j}$ moved to the left or to the right of ${\bf p}$. (right) The case where $b_jc$ and $b_{j+1}c'$ are chords, with ${\bf b_j}$ moved to the left or to the right of ${\bf p}$. In the first case, note that $\R_2$ contains the intersection point of ${\bf a}$ and ${\bf c'}$ inside the triangle, but it could be the case that the representation $\R_2$ obtained by induction has this intersection point located outside the triangle. See for example the illustration on the right part, where the intersection point of ${\bf b_j}$ and ${\bf c}$ is outside the triangle.}
    \label{fig:resolve-6pt-3-24}
\end{figure}

If $ab_{j+1}$ and $a'b_j$ are chords, we proceed similarly (see Figure~\ref{fig:resolve-6pt-4-24}). One difference is that, $T_1$ is the triangulation 
inside the cycle $(a,c,b_{j+1})$ and that $T_2$ is the one obtained from the
interior of the $5$-cycle $(b_{j+1},a',c',b_j,a)$
by adding the edges $ac'$ and $b_{j+1}c'$. Another difference is that we proceed differently according to the shape of $\R_2$. We distinguish the case where the intersection point of ${\bf a'}$ and ${\bf b_j}$ lies inside the triangle ${\bf ab_{j+1}c'}$ or outside it.

\begin{figure}
    \centering
    \includegraphics[width=\linewidth]{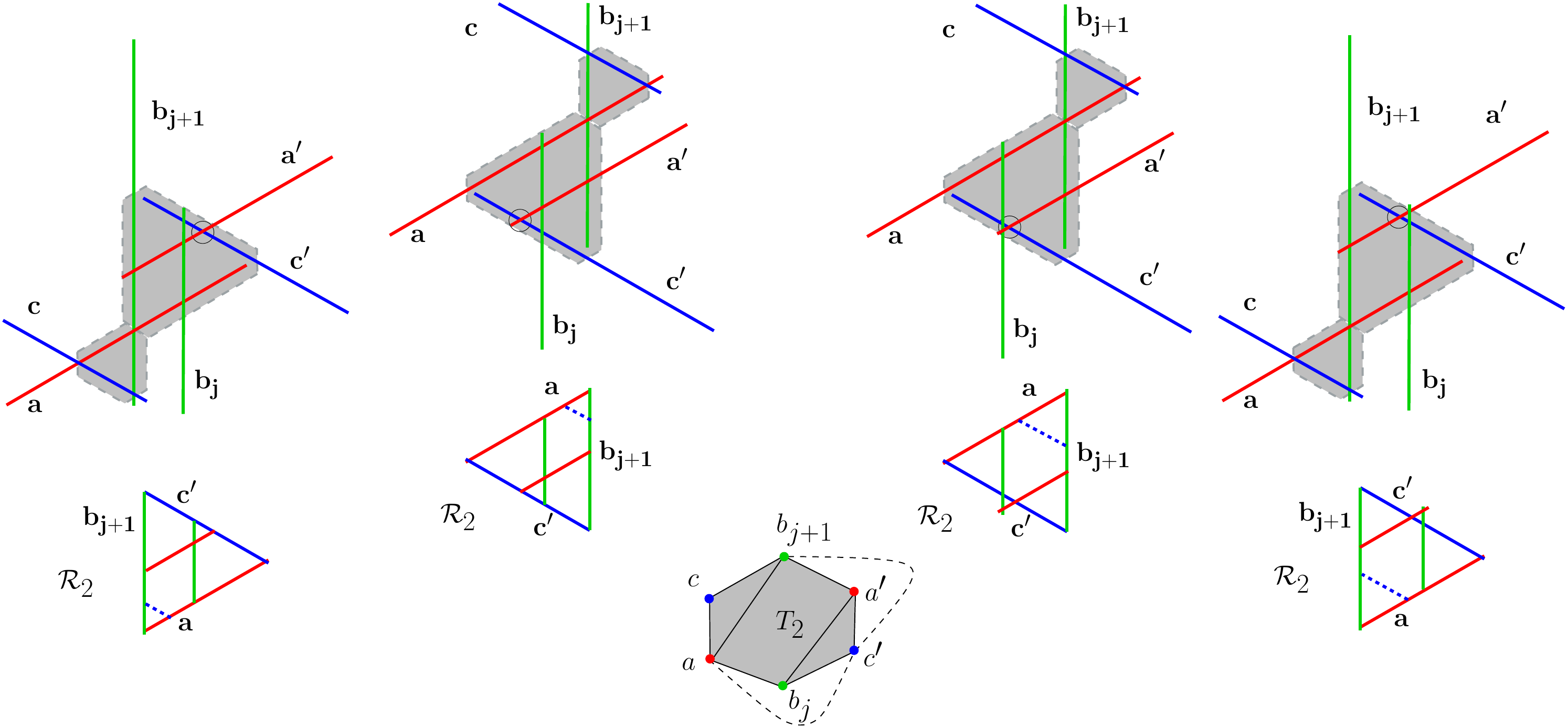}
    \caption{The case where $ab_{j+1}$ and $a'b_j$ are chords. (left) If in $\R_2$, the intersection point of ${\bf a'}$ and ${\bf b_j}$ is inside the triangle ${\bf ab_{j+1}c'}$. The two subcases whether ${\bf b_j}$ is moved to the left or to the right of ${\bf p}$. (right) If in $\R_2$, the intersection point of ${\bf a'}$ and ${\bf b_j}$ is outside the triangle ${\bf ab_{j+1}c'}$.}
    \label{fig:resolve-6pt-4-24}
\end{figure}

Finally, note that the moves of ${\bf b_j}$ and ${\bf b_{j+1}}$ are of proportional magnitudes (up to some constant depending
on the shapes of $\R'$ or $\R_2$). This property propagates from ${\bf b_1}$, to ${\bf b_t}$, and it is thus possible to assume that all these moves are arbitrarily small. The procedure hence leads to the sought 3-slope segment representation, and this
concludes the proof of the proposition.
\end{proof}

\section{Conclusion}
\label{sec:ccl}

Our result implies that for $k\le 3$, planar graphs that are
$k$-colorable admit a $k$-slopes segment representation, where
parallel segments induce an independent set. These graphs have a
so-called \emph{PURE-$k$-DIR} representation. Unfortunately, this does
not extend to the final case $k=4$ as conjectured by
D. West~\cite{W91}. The author~\cite{G19contrex} built a
counter-example based on a construction of Kardo\v{s} and
Narboni~\cite{KN19}. Their construction is an example of a signed
planar graph that is not 4-colorable, in the sense of signed graphs.
%, and it thus contradicts a conjecture of E.~Máčajová, A.~Raspaud, and M. Škoviera~\cite{MRS16}.
However, it remains open to know whether (4-colorable) planar graphs
admit a PURE-$k$-DIR representation (resp. a non-necessarily pure one)
for some $k>4$ (resp. $k>2$).

\paragraph{\bf Acknowledgements.}
The author is thankful to the anonymous reviewers, especially for a simplified proof
of Lemma~\ref{lem:det}, and for pointing out a gap in a preliminary version of the proof. The author is also thankful to Pascal Ochem for bringing~\cite{KN19} to
his attention, and to Marc de Visme for fruitful discussions on this
topic.

\end{document}